\documentclass[%
 aip,
 amsmath,amssymb,
 groupedaddress,
 reprint,%
]{revtex4-1}

\usepackage{graphicx}
\usepackage{dcolumn}
\usepackage{bm}

\usepackage[utf8]{inputenc}
\usepackage[T1]{fontenc}
\usepackage{mathptmx}
\usepackage{etoolbox}

\newcommand{\bq}{\boldsymbol{q}}
\newcommand{\bn}{\boldsymbol{n}}
\newcommand{\bj}{\boldsymbol{j}}
\newcommand{\br}{\boldsymbol{r}}
\newcommand{\bx}{\boldsymbol{x}}

\newcommand{\bz}{\boldsymbol{z}}

\def\bUtrap{\boldsymbol{U}^\mathrm{trap}}

\newcommand{\bA}{\boldsymbol{A}}

\newcommand{\bD}{\boldsymbol{D}}

\newcommand{\bF}{\boldsymbol{F}}

\newcommand{\bI}{\boldsymbol{I}}

\newcommand{\bL}{\boldsymbol{L}}
\newcommand{\bM}{\boldsymbol{M}}

\newcommand{\bQ}{\boldsymbol{Q} }

\newcommand{\bU}{\boldsymbol{U}}

\newcommand{\bOmega}{\bm{\varOmega}}

\def\bPhi{\bm{\Phi}}

\def \avg#1{\langle #1\rangle}
\def \bigavg#1{\bigl\langle #1\bigr\rangle}
\def \Bigavg#1{\Bigl\langle #1\Bigr\rangle}
\DeclareMathOperator{\Var}{Var}
\DeclareMathOperator{\Cov}{Cov}

\newcommand{\ksts}{k_sT_s}
\newcommand{\kt}{k_BT}

\usepackage{xcolor}

\makeatletter
\def\@email#1#2{%
 \endgroup
 \patchcmd{\titleblock@produce}
  {\frontmatter@RRAPformat}
  {\frontmatter@RRAPformat{\produce@RRAP{*#1\href{mailto:#2}{#2}}}\frontmatter@RRAPformat}
  {}{}
}%
\makeatother
\begin{document}


\title[Trapped-particle microrheology]{Trapped-particle microrheology of active suspensions}
\author{Zhiwei Peng}
\author{John F. Brady}%
 \email{jfbrady@caltech.edu}
\affiliation{ 
Division of Chemistry and Chemical Engineering, California Institute of Technology, Pasadena, California 91125, USA
}%

\date{\today}

\begin{abstract}
In microrheology, the local rheological properties such as viscoelasticity of a complex fluid are inferred from the free or forced motion of embedded colloidal probe particles. Theoretical machinery developed for forced-probe microrheology of colloidal suspensions focused on either constant-force (CF) or constant-velocity (CV) probes while in experiments neither the force nor the kinematics of the probe is fixed. More importantly, the constraint of CF or CV introduces a difficulty in the meaningful quantification of the fluctuations of the probe due to a thermodynamic uncertainty relation. It is known that for a Brownian particle trapped in a harmonic potential well, the product of the standard deviations of the trap force and the particle position is $d\kt$ in $d$ dimensions with $\kt$ being the thermal energy. As a result, if the force (position) is not allowed to fluctuate, the position (force) fluctuation becomes infinite. To allow the measurement of fluctuations, in this work we consider a microrheology model in which the embedded probe is dragged along by a moving harmonic potential so that both its position and the trap force are allowed to fluctuate. Starting from the full Smoluchowski equation governing the dynamics of $N$ hard active Brownian particles, we derive a pair Smoluchowski equation describing the dynamics of the probe as it interacts with one bath particle by neglecting hydrodynamic interactions among particles in the dilute limit. From this, we determine the mean and the variance (i.e., fluctuation) of the probe position in terms of the pair probability distribution. We then characterize the behavior of the system in the limits of both weak and strong trap. By taking appropriate limits, we show that our generalized model can be reduced to the well-studied CF or CV microrheology models.
\end{abstract}

\maketitle

\section{Introduction}
\label{sec:Introduction-trap}
Rheology is the study of flow and deformation of complex materials in response to an applied force. Traditional (bulk) rheological measurements are performed by shearing a macroscopic sample of the material confined between two solid surfaces, such as in the cone-and-plate rheometer. Bulk rheological studies such as shear rheometry provide a measurement of the macroscopic rheological behavior of complex materials.

Recently, particle-tracking microrheology has become a standard tool for studying the mechanical properties of materials on a much smaller scale \citep{Weihs2006,cicuta2007microrheology,wirtz2009particle,furst2017microrheology}. In contrast to bulk rheology,  microrheology only requires a small sample volume and can be used to quantify spatial heterogeneity. As a result, microrheology 
is particularly useful for examining soft biological materials. For example, classical bulk rheometry cannot be used to probe the microenvironment inside living cells without disrupting their mechanical structure while particle-tracking microrheology can be  performed  \citep{Wilhelm2003,Nawaz2012,berret2016local,ayala2016rheological,Hu2017}.

To aid in the understanding of experimental measurements and in the prediction of colloidal microrheology, \citet{SquiresBrady2005} developed a theoretical framework in which a colloidal probe is pulled through a suspension of neutrally buoyant bath colloids. This model has been used and generalized to study the microrheology of passive colloids \citep{Khair2005viscoelastic,khair_brady_2006,meyer2006,zia_brady_2010,swan2013,Zia2018} and active colloids \citep{Burkholder2019JCP,Burkholder2020}. When the external pulling force is absent, the probe  ``collides'' with bath particles as it undergoes Brownian motion---the so-called tracer diffusion problem. To characterize the nonlinear response, forced microrheology is considered in which an external force, often larger than the thermodynamic restoring force, is applied to the probe. Within forced microrheology, two  operating modes---constant-force (CF) and constant-velocity (CV)---are often considered from a theoretical perspective. In the CF mode, the probe is driven by a constant external force $\bF^\text{ext}$ and the velocity of the probe is fluctuating. Conversely, for a CV probe, the probe velocity $\bU_1$ is a constant vector (Therefore, the position of the probe is known at all times.) and the force required to maintain such a steady motion must fluctuate. 

To characterize the micro-viscous response of colloidal suspensions, an effective microviscosity $\eta^\text{eff}$ can be defined using the Stokes drag law. For a spherical probe of radius $a$ in the CF mode,  this is given by $F^\text{ext} = 6 \pi \eta^\text{eff} a\langle U_1 \rangle$, where $\langle U_1 \rangle$ the probe velocity in the direction of $\bF^\text{ext}$ averaged over Brownian fluctuations. The ratio between the effective microviscosity and the solvent viscosity, $\eta^\text{eff}/\eta$, is the main quantity of interest in colloidal microrheology.  For the CV mode, the average external force is used in the definition of the effective microviscosity: $\langle F^\text{ext} \rangle = 6 \pi \eta^\text{eff} aU_1$. In order to measure the microviscoelastic response of suspensions, an oscillatory driving force is considered \citep{Khair2005viscoelastic}.

While the CF (or CV) model is successful in quantifying the mean velocity (or mean force) of a probe driven through colloidal suspensions. The fluctuation from this mean value is largely unexplored. Taking the CV mode as an example, one could calculate the variance of the mean force using the probe-distorted microstructure. The question is what does this variance physically imply? In particular, how does this variance relate to the fluctuations in the suspension? In an experimental setting, neither the force nor the velocity of the probe is fixed; they are both allowed to fluctuate \citep{meyer2006,Weihs2006,cicuta2007microrheology, yao2009microrheology}.

To mimic the experimental realization more closely and motivate later discussions, consider the simple case of an isolated Brownian particle in a harmonic trap that is centered at the origin (arbitrary). In this physical picture, both the position and the velocity of the particle is fluctuating. A statistical mechanical description can be adopted in which one defines the probability density, $P(\br, t)$, of finding the particle at position $\br$ relative to the fixed trap at time $t$. Conservation of probability dictates that $P(\br, t)$ is governed by the Smoluchowski equation, which reads $\partial P/\partial t + \nabla\cdot\bj =0$, where the flux vector $\bj = P\bF^\text{trap}/\zeta - D_T \nabla P$. Here, $\bF^\text{trap}$ is the trap force and for a harmonic trap is given by $\bF^\text{trap} = -k \br$ with $k$ being the spring constant; $\zeta$ is the drag coefficient and $D_T$ is the thermal diffusivity given by the Stokes-Einstein-Sutherland relation, $\zeta D_T = k_BT$, where $k_BT$ is the thermal energy. The mean external force exerted on the Brownian particle is $\langle \bF^\text{trap} \rangle = \int \bF^\text{trap}  P d\br = -k \int \br P d\br = -k \langle \br \rangle $. Because the trap is harmonic, the mean force is proportional to the mean displacement with $-k$ being the constant of proportionality. For a fixed trap, the mean position (therefore the mean force) is zero, $\langle \br \rangle = \bm{0}$. The variance of the force, $\Var\left(\bF^\text{trap} \right) = k^2 \Var(\br)$. A straightforward calculation leads to the result
\begin{equation}
\label{eq:Brownian-mean-position}
    \Var(\br) = \frac{k_BT}{k}\bI, 
\end{equation}
where $\bI$ is the identity tensor. Introducing the shorthand $\Delta \bF^\text{trap} = \bF^\text{trap} - \langle  \bF^\text{trap}\rangle $, we can write the fluctuation relation as 
\begin{equation}
\label{eq:Brownian-dkBT}
    \left< (\Delta  \bF^\text{trap})^2 \right>^{1/2}\left< (\Delta  \br_1)^2 \right>^{1/2} = d \kt,
\end{equation}
where $d$ is the spatial dimensionality.


Equation \eqref{eq:Brownian-dkBT} is a fundamental result and a few comments on its implications are in order. First, by harmonically trapping a particle immersed in a solvent, the product of the standard deviations of the trap force and the particle position gives precisely the thermal fluctuations of the solvent---$dk_BT$. Second, one can decrease the uncertainty in the position by increasing the stiffness of the trap [see equation \eqref{eq:Brownian-mean-position}]. However, the trade-off is that the fluctuation in the force must increase due to \eqref{eq:Brownian-dkBT}. Said differently, this constitutes a thermodynamic uncertainty relation in which one cannot decrease the fluctuations in both the force and the position simultaneously. If the fluctuation in the position vanishes (infinitely stiff trap), the fluctuation in the force blows up.

We note that \eqref{eq:Brownian-dkBT} is observed elsewhere. For example, consider an ideal Gaussian polymer chain with one end localized in a harmonic trap. The fluctuations of the trap force and the position from the trap center satisfy an identical relation \citep{wang201750th}.

We are now in a position to consider the fluctuations in the microrheology problem. Instead of considering either CF or CV, we must allow both the position of and the force on the probe to fluctuate in order to have a meaningful quantification of fluctuations. Equation \eqref{eq:Brownian-dkBT} also implies that we should consider the position not the velocity of the probe. In the CF mode, therefore, the quantity of interest for fluctuations is the variance of the position of the probe, which is just the force-induced tracer diffusion problem. That is, the tracer diffusivity under the influence of a constant force should be considered---not the variance of the velocity. For the CV mode, the position of the probe is also prescribed and the fluctuation in the force is infinite. As a result, in the CV mode the computed variance of the force does not have a physical meaning.

In this paper, to closely mimic the setup of microrheological experiments, we consider a trapped-particle microrheology model in which the colloidal probe particle is driven by a translating harmonic trap. Because biological materials examined by microrheology such as the microenvironment inside living cells often contain active ``particles'', we model the suspension as an active colloidal suspension. Compared to passive suspensions, the study of the microrheology of active suspensions is more recent \citep{jepson13,Clement13,Morozov14,Kasyap14,Reichhardt,Burkholder2017,Burkholder2019JCP,Burkholder2020,knevzevic2021oscillatory,peng2022forced,Ahmed22}. The colloidal particles in an active suspension are able to self-propel, which can be a model for either biologically active microswimmers or synthetic phoretic particles.  This active colloidal suspension model also includes passive (not self-propelled) colloidal systems, which can be obtained by setting the self-propulsive swim speed to zero.

The paper is organized as follows. In section \ref{sec:active-suspension-mechanics}, we present the general $N$-particle dynamics from a continuum perspective using the Smoluchowski equation governing the evolution of the positions and orientations of $N$ active Brownian particles. In section \ref{sec:moving-trap-microrheology} we first derive the mean and variance (fluctuation) of the probe position relative to the trap center from the $N$-particle formulation. Neglecting hydrodynamic interactions in the dilute limit, we then derive the pair-level Smoluchowski equation governing the dynamics of the probe and one bath particle. We discuss the asymptotic behavior of the system in the limits of both weak and strong traps.  We then show in section \ref{sec:CF-and-CV} that  our generalized theoretical framework includes the well-studied CF and CV microrheology models when appropriate limits are taken. Finally, we conclude in section \ref{sec:conclusion}.

\section{Mechanics of active Brownian suspensions}
\label{sec:active-suspension-mechanics}

Consider a colloidal suspension consisting of $N$ particles dispersed in an incompressible Newtonian fluid (solvent) of dynamic viscosity $\eta$. The particles could be active and are subject to fluctuating thermal (Brownian) forces from the solvent. Furthermore, the inertia of the fluid and the particles are assumed to be negligible. In this low-Reynolds-number regime, the fluid dynamics is governed by the linear Stokes equations and the probability distribution of the particles are described by the Smoluchowski equation. In general, all $N$ particles could be active, and we model them as active Brownian particles. The probability distribution for finding the $N$ particles in  positions $\{\bx_\alpha\}$ and orientations $\{\bq_\alpha\}$  at a given time $t$ is denoted as $P_N(\bx^N, \bq^N, t)$ where $\alpha = 1,\cdot\cdot\cdot,N$ is the particle label. In the laboratory frame of reference, the $N$-particle Smoluchowski equation is given by
\begin{equation}
    \label{eq:N-particle-general-Smoluchowski}
    \frac{\partial P_N}{\partial t} + \sum_{\alpha=1}^N\nabla_\alpha^T \cdot\bj_\alpha^T + \sum_{\alpha=1}^N \nabla_\alpha^R \cdot\bj_\alpha^R = 0,
\end{equation}
where $\nabla_\alpha^T = \partial/\partial \bx_\alpha$ is the spatial gradient operator with respect to the position vector ($\bx_\alpha$) of particle $\alpha$ in the laboratory frame and $\nabla_\alpha^R = \bq_\alpha\times \left( \partial/\partial \bq_\alpha\right)$ is the orientational gradient operator of particle $\alpha$. The translational and rotational fluxes in equation \eqref{eq:N-particle-general-Smoluchowski} are, respectively, given by $\bj_\alpha^T = \bU_\alpha P_N$ and $\bj_\alpha^R = \bOmega_\alpha P_N$, where $\bU_\alpha$ ($\bOmega_\alpha$) is the instantaneous linear (angular) velocity  of particle labeled $\alpha$ relative to the laboratory frame. The conservation of probability is
\begin{equation}
    \label{eq:PN-conservation}
    \int_{\Gamma_N} P_N  d\Gamma^N = 1,
\end{equation}
where $d\Gamma^N = \prod_{\alpha=1}^N d \Gamma_\alpha$ denotes the volume element of the $N$-particle phase space and $d \Gamma_\alpha = d \bx_\alpha d \bq_\alpha$ is the volume element of the phase space of particle $\alpha$.

In the absence of a background flow, the linear and angular velocities of any active particle $\alpha$ are given by
\begin{align}
    \label{eq:mobility-relation-general}
    \begin{pmatrix}
        \bU_\alpha - \bU_\alpha^0 \\
        \bOmega_\alpha - \bOmega_\alpha^0
    \end{pmatrix}
    = &\sum_{\beta=1}^N \bm{\mathcal{M}}_{\alpha\beta} \cdot
    \begin{pmatrix}
        \bF_\beta^e + \bF_\beta^P - \kt \nabla_\beta^T \ln P_N \\
        \bL_\beta^e + \bL_\beta^P - \kt \nabla_\beta^R \ln P_N
    \end{pmatrix} \nonumber \\
    &+\begin{pmatrix}
        \bm{0}\\
        - D_\alpha^R \nabla_\alpha^R \ln P_N
    \end{pmatrix},
\end{align}
where $\bm{\mathcal{M}}_{\alpha\beta}$ is the configuration-dependent grand hydrodynamic mobility tensor coupling the linear and angular velocity of particle $\alpha$ to the force and torque exerted on particle $\beta$. Note that for general particle shapes $\bm{\mathcal{M}}_{\alpha\beta}$ is a function of the instantaneous $N$-particle configuration---both positions and orientations. The forces on any particle $\beta$ include the external force $\bF_\beta^e$, the interparticle colloidal force $\bF_\beta^P$ and the thermal or entropic force $-\kt\nabla_\beta^T\ln P_N$. Similarly, the torques on any particle $\beta$ include the external torque $\bL_\beta^e$, the interparticle colloidal torque $\bL_\beta^P$ and the thermal torque $-\kt\nabla_\beta^R\ln P_N$. The interparticle colloidal forces and torques are assumed to be conservative. For the case of hard-sphere interactions, the interparticle forces reduce to no-flux boundary conditions at any surface of contact between particles.

In equation \eqref{eq:mobility-relation-general}, the activity of any particle $\alpha$ is modeled by its undisturbed swim linear velocity $\bU_\alpha^0$ and angular velocity $\bOmega_\alpha^0$ regardless of the presence of any other particles. For the case of simple ABPs, the swim angular velocity  is  often taken to be zero, $\bOmega_\alpha^0 = \bm{0}$. Furthermore, a biological microswimmer may ``decide'' to change its orientation $\bq_\alpha$ by, for example, actuating the flagella on a different side of its body without disturbing the flow. In this process, the body of the microswimmer does not turn. For non-spherical particles, this process means that the swim orientation $\bq_\alpha$ is usually different from the orientation of the particle shape, in which case the shape orientation needs to be included as an additional phase space variable. For spherical particles, only the swim orientation matters and no such difficulty is introduced. This reorientation process of any particle $\alpha$ is independent of the motion of other particles and is modeled by a simple rotary diffusion with a constant rotary diffusivity $D_\alpha^R$. The reorientation time is $\tau_\alpha^R = 1/D_\alpha^R$, which defines the active run or persistence length of an ABP: $\ell_\alpha = U_\alpha^0 \tau_\alpha^R$. Because this reorientation process is biological rather than thermal in origin, $D_\alpha^R$ is not constrained by the fluctuation-dissipation theorem and may be inferred from experimental data.

\section{Moving-trap microrheology}
\label{sec:moving-trap-microrheology}
In the context of microrheology, the particle with label $1$ is identified as the probe particle.  This particle could be a new particle placed into the suspension or one of the suspension particles tagged as the probe. Particles labeled $2-N$ are referred to as bath particles. In the following, we consider a suspension of neutrally buoyant, hard and active colloidal spheres with identical radii. The probe may have a different radius than the bath particles. Instead of fixing the external force $\bF_1^e$ or the velocity $\bU_1$, the probe particle is trapped in a translating harmonic potential well. Denoting the position vector of the center of the potential well as $\bx_0(t)$, we have $d\bx_0/dt = \bUtrap(t)$, where $\bUtrap(t)$ is the prescribed velocity of the moving trap relative to the laboratory frame. The trap force $\bF_1^e$ is assumed to be only a function of the relative position between the probe and the potential well. All bath particles experience no external forces or torques. We first consider a general derivation in which all particles are ABPs and the probe is a tagged ABP in the suspension.

In the constant-force or constant-velocity mode of microrheology, the position of the probe does not matter, and the system is statistically homogeneous. In contrast, the introduction of a moving trap defines a specific origin in the system and the position of the probe relative to the trap needs to be considered explicitly. To this end, we first change to a coordinate system moving with the instantaneous trap velocity and measure all particle positions relative to the trap. This change of variables is written as $\bz_\alpha = \bz_\alpha(\{\bx\}, t) = \bx_\alpha - \int_0^t \bUtrap(s)ds - \bx_0(0)$ for any $\alpha$ and $t^\prime = t^\prime(\{\bx\}, t) = t$. Using the chain rule we obtain $\partial/\partial t = - \sum_{\alpha=1}^N \bUtrap\cdot\partial/\partial \bz_\alpha +\partial/\partial t^\prime$ and $\partial/\partial\bx_\alpha = \partial/\partial \bz_\alpha$. The Smoluchowski equation \eqref{eq:N-particle-general-Smoluchowski} in the new coordinate system becomes
\begin{equation}
    \label{eq:smoluchowski-movingtrap-z-coordinate}
    \frac{\partial P_N}{\partial t^\prime } + \sum_{\alpha=1}^N \frac{\partial}{\partial \bz_\alpha}\cdot\left(\bj_\alpha^T - \bUtrap P_N\right) + \sum_{\alpha=1}^N \nabla_\alpha^R\cdot\bj_\alpha^R=0,
\end{equation}
where $\bj_\alpha^T$ and $\bj_\alpha^R$ remain unchanged. In the context of microrheology, it is more convenient to measure the positions of all bath particles relative to that of the probe. We therefore introduce another change of variables such that for the probe $\br_1 = \br_1(\bz^N, t^\prime) = \bz_1 $, and $\br_\alpha =\br_\alpha(\bz^N, t^\prime) = \bz_\alpha - \bz_1 $ for all bath particles ($\alpha = 2,\cdot\cdot\cdot,N$). In this coordinate system, the probe position is measured relative to the trap and the positions of all bath particles are measured relative to the probe. The change of variables allows us to write $\partial/\partial \bz_1 = \partial/\partial \br_1 - \sum_{\alpha=2}^N \partial/\partial\br_\alpha$ and $\partial/\partial\bz_\alpha = \partial/\partial\br_\alpha $ for $\alpha = 2,\cdot\cdot\cdot,N$. The Smoluchowski equation \eqref{eq:smoluchowski-movingtrap-z-coordinate} transforms to
\begin{align}
    \label{eq:smoluchowski-movingtrap-r-coordinate}
    \frac{\partial P_N}{\partial t } + \nabla_1^T\cdot\left(\bj_1^T - \bUtrap P_N\right) &+ \sum_{\alpha=2}^N \nabla_\alpha^T\cdot\left(\bj_\alpha^T - \bj_1^T\right) \nonumber \\
    &+ \sum_{\alpha=1}^N \nabla_\alpha^R\cdot\bj_\alpha^R=0.
\end{align}

It is understood that in equation \eqref{eq:smoluchowski-movingtrap-r-coordinate} we have used $t$ for the time variable and $\nabla_\alpha^T = \partial/\partial\br_\alpha$ for any $\alpha$. Formally, the probability density in equation \eqref{eq:smoluchowski-movingtrap-r-coordinate} is the conditional probability of find all particles at a given configuration provided that the trap is at $\bx_0$ at time $t$, i.e., $P_N = P_N\left(\br^N, \bq^N, t\rvert \bx_0, t\right)$. The translational flux of particle $\alpha$ can be written as
\begin{eqnarray}
    \label{eq:N-particle-translational-flux}
    \bj_\alpha^T &=& U_\alpha^0 \bq_\alpha P_N +  \bM_{\alpha 1}^{UF}\cdot\bF_1^eP_N \nonumber \\
    & &- \sum_{\beta=1}^N \left(\bD_{\alpha\beta}^{UF} - \bD_{\alpha 1}^{UF}\right)\cdot\nabla_\beta^T P_N\nonumber\\
    & & - \bD_{\alpha1}^{UF}\cdot\nabla_1^T P_N - \sum_{\beta=1}^N \bD_{\alpha\beta}^{UL} \cdot\nabla_\beta^R P_N,
\end{eqnarray}
where we have taken $\bU_\alpha^0 = U_\alpha^0 \bq_\alpha$ and used the Stokes-Einstein-Sutherland relations $\bD_{\alpha\beta}^{UF} = \kt \bM_{\alpha\beta}^{UF}, \bD_{\alpha\beta}^{UL} = \kt \bM_{\alpha\beta}^{UL}$. For all accessible configurations, the inter-particle forces are zero and the hard-particle interaction between two spheres do not induce torques. Similarly, the rotary flux of particle $\alpha$ is given by
\begin{eqnarray}
    \label{eq:N-particle-rotational-flux}
    \bj_\alpha^R &= & \bM_{\alpha 1}^{\Omega F}\cdot\bF_1^eP_N - \sum_{\beta=1}^N \left(\bD_{\alpha\beta}^{\Omega F} - \bD_{\alpha 1}^{\Omega F}\right)\cdot\nabla_\beta^T P_N  \nonumber \\
    &&- \bD_{\alpha1}^{\Omega F}\cdot\nabla_1^T P_N - \sum_{\beta=1}^N \bD_{\alpha\beta}^{\Omega L} \cdot\nabla_\beta^R P_N -D_\alpha^R \nabla_\alpha^R P_N.
\end{eqnarray}
There are no external force or torque on the bath particles, $\alpha=2\mbox{--}N$, nor a torque on the probe, $\bL_1^e = \bm{0}$.

The Smoluchowski equation \eqref{eq:smoluchowski-movingtrap-r-coordinate} together with the flux expressions \eqref{eq:N-particle-translational-flux} and \eqref{eq:N-particle-rotational-flux} fully specify the $N$-particle phase space dynamics. Some comments regarding equations \eqref{eq:smoluchowski-movingtrap-r-coordinate}-\eqref{eq:N-particle-rotational-flux} are in order. First, the above derivation is an extension of the model considered by  \citet{SquiresBrady2005} for passive Brownian suspensions. We have generalized their model to a suspension of ABPs in which one of the particles is tagged as the probe that is driven by a translating trap. Realizing that the grand mobility tensor does not depend on the swim orientation vectors of spherical particles,  one can set $\bU_\alpha^0 =0$ and integrate over the orientations of all particles to obtain the trapped probe microrheology problem of a passive Brownian suspension. Note that even for passive suspensions, if the probe or the bath particles are non-spherical, their shape orientations need to be included in the above formulation. Second, the hydrodynamic interactions between all $N$-particles are included in the grand mobility tensor. In particular, this leads to the fact that a gradient in orientation space of particle $\beta$ induces a translational flux of particle $\alpha$, and vice versa,  due to the hydrodynamic translation-rotation coupling. Third, due to the dependence on particle orientations, the phase space of $N$ ABPs has a dimension of $5N$: the physical space has a dimension of $3N$ and the orientation space has a dimension of $2N$ if the orientation of each particle is parametrized by the azimuthal and polar angles of a spherical coordinate system.

\subsection{Mean and fluctuation of the probe position}
The average position or mean displacement of the probe relative to the trap is defined by
\begin{equation}
    \label{eq:avg-PN}
    \avg{\br_1}(t) = \int \br_1 P_N d\Gamma^N,
\end{equation}
where the angle bracket denotes integration against $P_N$ over the configuration space of all particles. Multiplying equation \eqref{eq:smoluchowski-movingtrap-r-coordinate} by $\br_1$ and integrating over the configuration space $\Gamma^N$, we obtain
\begin{eqnarray}
    \label{eq:md-PN}
    \frac{\partial \avg{\br_1} }{\partial t} +\bUtrap  &= & U_1^0 \avg{\bq_1} + \bigavg{\bM_{11}^{UF}\cdot\bF_1^e}
    - \bigavg{ \bD_{11}^{UF}\cdot\nabla_1^T\ln P_N} \nonumber \\
   & & - \sum_{\beta=1}^N \bigavg{ \left(\bD_{1\beta}^{UF} - \bD_{1 1}^{UF}\right)\cdot \nabla_\beta^T \ln P_N }.
\end{eqnarray}
Similarly, the mean squared displacement, a second order tensor,  is governed by
\begin{equation}
    \label{eq:msd-PN}
    \frac{\partial \avg{\br_1\br_1}}{\partial t} + 2\left[\bUtrap\avg{\br_1}\right]^\mathrm{sym}  =  2 \int \left[ \bj_1^T \br_1\right]^\mathrm{sym}d\Gamma^N ,
\end{equation}
where the integral
\begin{eqnarray}
    \int \bj_1^T\br_1 d \Gamma^N&= & U_1^0 \avg{\bq_1\br_1} + \Bigavg{\bM_{11}^{UF}\cdot\bF_1^e \br_1}\nonumber \\
    && - \sum_{\beta=1}^N \Bigavg{ \left(\bD_{1\beta}^{UF} - \bD_{1 1}^{UF}\right)\cdot \left(\nabla_\beta^T \ln P_N\right) \br_1 }\nonumber\\
    &&- \Bigavg{ \bD_{11}^{UF}\cdot\left(\nabla_1^T\ln P_N\right) \br_1} ,
\end{eqnarray}
and the superscript ``sym'' denotes the symmetric part of a tensor (see equation \eqref{eq:symmetric-tensor-def}).

The main quantities of interest in the present problem are the mean displacement $\avg{\br_1}$ and the fluctuation
\begin{align}
    \Var(\br_1)&=\Cov(\br_1, \br_1)=\bigavg{\Delta\br_1\Delta\br_1} \nonumber\\
    &= \bigavg{\br_1\br_1} - \avg{\br_1}\avg{\br_1},
\end{align}
where we have introduced the shorthand $\Delta \br_1 = \br_1 - \avg{\br_1}$ and $\Var(\br_1)$ denotes the variance tensor of $\br_1$.  For a harmonic trap, the mean force is related to the mean displacement via 
\begin{equation}
    \avg{\bF_1^e} = -k \avg{\br_1},
\end{equation}
 and similarly the fluctuation in the force is given by 
 \begin{equation}
    \Var(\bF_1^e)=\bigavg{\Delta\bF_1^e \Delta\bF_1^e} = k^2 \avg{\Delta\br_1\Delta\br_1}.
 \end{equation}

\subsection{The pair problem}
To proceed analytically, we restrict the analysis to the dilute limit in which only pair interactions between a bath particle and the probe is considered. Furthermore, we neglect hydrodynamic interactions between the bath particle and the probe,  and only consider hard-sphere interactions. The reduction from the $N$-particle formulation to the pair problem and the consideration of hydrodynamic interactions are discussed in appendix \ref{sec:pair-derivation}.

Because the bath particles are indistinguishable, it is convenient to define the two-particle probability density function $\rho_2(\br_2, \bq_2, \br_1, \bq_1, t)$, which denotes the joint probability density function of finding the probe at $(\br_1, \bq_1)$ and any bath particle at  $(\br_2, \bq_2)$ at time $t$. In terms of $P_2(\br_2, \bq_2, \br_1, \bq_1, t)$, which is the joint probability density function of finding the probe at $(\br_1, \bq_1)$ and the bath particle labeled $2$ (i.e., the first bath particle) at  $(\br_2, \bq_2)$ at time $t$, we have $\rho_2 = (N-1)P_2$. Here, the factor of $N-1$ comes from removing the labels from the $N-1$ bath particles. The joint probability can be written as 
\begin{align}
    \rho_2 &= \rho_{1/1}(\br_2, \bq_2, t \rvert \br_1, \bq_1, t) P_1(\br_1, \bq_1, t) \nonumber \\
    &= n_b g_{1/1}(\br_2, \bq_2, t \rvert \br_1, \bq_1, t) P_1(\br_1, \bq_1, t),
\end{align}
where $n_b=(N-1)/V$ is the number density of bath particles. For a passive and CF (or CV) probe, $g_{1/1}$ becomes independent of the configuration ($\br_1$ and $\bq_1$) of the probe due to statistical homogeneity; in this case the probe distribution $P_1$ can be integrated over and one only needs to consider $g_{1/1}$ \citep{SquiresBrady2005,burkholder2018,Burkholder2019JCP}.

The joint probability $\rho_2$  (see appendix \ref{sec:pair-derivation}) is governed by
\begin{align}
    \label{eq:smoluchowski-pair-noHI}
    \frac{\partial \rho_2}{\partial t  } &+\nabla_1^T\cdot\left( \bj_1^T -\bUtrap \rho_2\right) + \nabla_2^T\cdot \left( \bj_2^T -\bj_1^T\right) \nonumber\\
    &+  \sum_{\alpha=1}^2 \nabla_\alpha^R\cdot\bj_\alpha^R=0,
\end{align}
where 
\begin{align}
    \label{eq:j1T-noHI}
    & \bj_1^T = U_1^0 \bq_1 \rho_2 + \frac{1}{\zeta_1}\bF_1^e \rho_2 + D_1^T  \nabla_2^T \rho_2 - D_1^T \nabla_1^T \rho_2,\\
    \label{eq:j2t-noHI}
    & \bj_2^T = U_2^0 \bq_2 \rho_2 - D_2^T \nabla_2^T \rho_2,\\
    \label{eq:jR-noHI}
    & \bj_\alpha^R = -D_\alpha^R \nabla_\alpha^R \rho_2.
\end{align}

\begin{figure*}
    \centering
    \includegraphics[width=0.5\textwidth]{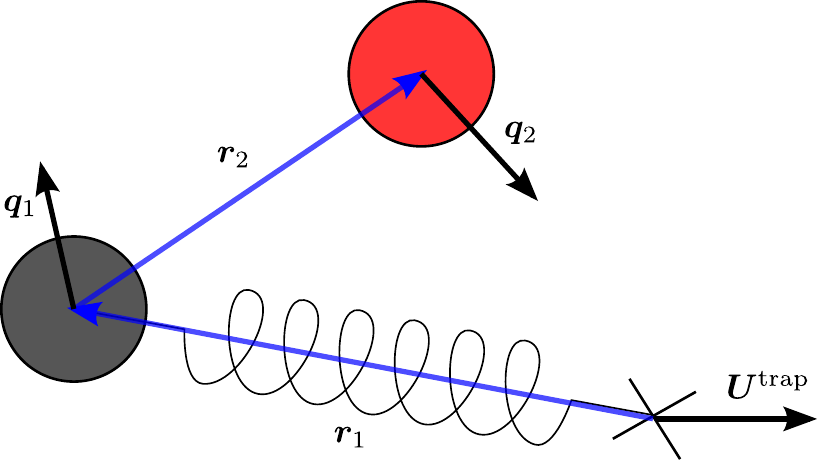}
    \caption{\label{fig:problem-schematic}Schematic of the pair problem of a spherical probe particle in a moving harmonic trap interacting with a spherical bath particle. Both the probe and the bath particles can be active.}
\end{figure*}

At contact, $r_2 = R_c$, no relative flux is allowed:
\begin{equation}
    \bn_2\cdot\left(\bj_2^T-\bj_1^T\right)=0.
\end{equation}
Far away from the probe, the bath distribution is undisturbed by the probe and the probe distribution is that in the absence of the bath particles,
\begin{equation}
    \label{eq:g2-r2-infinity-BC}
    \rho_2(\br_2, \bq_2, \br_1, \bq_1, t) \to \frac{n_b}{\Omega_b} P_1(\br_1,\bq_1, t)\quad \mathrm{as}\quad \lvert\br_2\rvert \to \infty,
\end{equation}
where $\Omega_b$ is the total solid angle of the orientation space of the bath particle. In 3D, $\Omega_b=4\pi$. Far away from the trap, the probability vanishes
\begin{equation}
    \label{eq:g2-r1-infinity-BC}
    \rho_2 \to 0 \quad\mathrm{as}\quad \lvert\br_1\rvert \to \infty.
\end{equation}

Equation \eqref{eq:md-PN} governing the mean displacement becomes
\begin{align}
    \label{eq:md-pair}
    \frac{\partial \avg{\br_1} }{\partial t}  + \frac{1}{\tau_k}\avg{\br_1} =&-\bUtrap+ U_1^0\avg{\bq_1}  \nonumber\\
    &+ D_1^T\int \nabla_2^T \rho_2 d\Gamma^2,
\end{align}
where $d\Gamma^2=d\Gamma_1d\Gamma_2$, and we have defined the viscoelastic timescale
\begin{equation}
 \tau_k = \frac{\zeta_1}{k},
\end{equation}
which is set by the balance between the viscous force $\zeta_1 \partial \avg{\br_1}/\partial t$ and the elastic force $k\avg{\br_1}$. Using the divergence theorem and the far-field condition \eqref{eq:g2-r2-infinity-BC}, the last term on the rhs of \eqref{eq:md-pair} can be written as
\begin{equation}
    D_1^T\int \nabla_2^T \rho_2 d\Gamma^2 = D_1^T\int d\bq_2 d\Gamma_1\oint_{S_c} \bn_2 \rho_2 dS_2,
\end{equation}
where $S_c=\{\br_2: \lvert\br_2\rvert=R_c \}$ is the contact surface and $\bn_2$ is the unit normal vector of $S_c$ that points out of particle $2$.

As shown in appendix \ref{sec:appendix-pair-covariance}, the position fluctuation of the probe is governed by
\begin{widetext}
\begin{equation}
        \label{eq:var-r1-q1r1-text}
        \frac{1}{2}\frac{\partial \Var(\br_1)}{\partial t}  + \frac{1}{\tau_k} \Var(\br_1)= D_1^T\bI +\left[ U_1^0\Cov(\bq_1, \br_1)+D_1^T\int \Delta\br_1\nabla_2^T\rho_2  d\Gamma^2  \right]^\mathrm{sym},
\end{equation}
where the covariance of $\bq_1$ and $\br_1$ satisfies
\begin{align}
    \label{eq:Covariance-q1-r1-pair}
    \frac{\partial \Cov(\bq_1, \br_1)}{\partial t} +\frac{1}{\tau} \Cov(\bq_1, \br_1)  =  U_1^0\Var(\bq_1) 
     + D_1^T\int \Delta \bq_1 \nabla_2^T\rho_2 d\Gamma^2.
\end{align}
\end{widetext}
In equation \eqref{eq:Covariance-q1-r1-pair}, we have defined the relaxation time $\tau$ using 
\begin{align}
    \frac{1}{\tau} =  \frac{1}{\tau_k}+\frac{d-1}{\tau_1^R}.
\end{align}
Regardless of the presence of the trap or the bath particles, at long times ($t\to\infty$) the net polar and nematic orders of the probe are given by $\avg{\bq_1}=\bm{0}$ and $\bigavg{\bq_1\bq_1} = \bI/d$, respectively (see appendix \ref{sec:appendix-pair-covariance}). As a result, $\Var(\bq_1) = \bI/d$ at long times.

It is convenient to consider the rank $m$ polyadic spatial moment tensor
\begin{equation}
    \bM_m(\br_2, \bq_2, \bq_1, t) = \int \underbrace{\br_1\cdot\cdot\cdot\br_1}_{m} \rho_2 d\br_1,\quad (m=0,1,2,...).
\end{equation}
Multiplying equation \eqref{eq:smoluchowski-pair-noHI} by the $m$-adic product of $\br_1$ and integrating over the physical space of the probe, we obtain
\begin{widetext}
\begin{eqnarray}
    &&\frac{\partial \bM_m}{\partial t} - m \left[U_1^0\bq_1 \bM_{m-1} - \frac{k}{\zeta_1}\bM_m +(m-1) D_1^T \bM_{m-2}\bI +D_1^T \nabla_2^T\bM_{m-1} - \bUtrap\bM_{m-1} \right]^\mathrm{sym}\nonumber \\
    && +\nabla_2^T\cdot\left(\bU_r^0 \bM_m - D_r^T \nabla_2^T\bM_m + \frac{k}{\zeta_1}\bM_{m+1}\right) - m D_1^T \left[\nabla_2^T \bM_{m-1} \right]^\mathrm{sym} - \sum_{\alpha=1}^2 D_\alpha^R\nabla_\alpha^R\cdot\nabla_\alpha^R \bM_m=\bm{0},
    \label{eq:equation-spatial-moment-P2}
\end{eqnarray}
\end{widetext}
where we have defined the relative swim velocity and the relative diffusivity as, respectively,
\begin{equation}
    \bU_r^0 = U_2^0\bq_2-U_1^0\bq_1,\quad D_r^T = D_1^T+D_2^T,
\end{equation}
whereas $\left[\bA\right]^\mathrm{sym}$ denotes the symmetric part of any rank $m$ Cartesian tensor $\bA$ such that
\begin{equation}
    \label{eq:symmetric-tensor-def}
    \left[\bA\right]^\mathrm{sym}_{i_1i_2\cdot\cdot\cdot i_m} =
    \frac{1}{m!} \sum_{\sigma \in \mathfrak{S}_m} A_{i_{\sigma 1} i_{\sigma 2}\cdot\cdot\cdot i_{\sigma m}},
\end{equation}
in which $\mathfrak{S}_m$ is the set containing the $m!$ permutations of indices. For $m=2$, this reduces to the familiar definition of the symmetric part of a rank $2$ tensor, $\bA^\mathrm{sym} = \left(\bA+\bA^\intercal \right)/2$. For any rank $m$ tensor $\bA$, its symmetric part $\left[\bA\right]^\mathrm{sym}$ is invariant under a permutation of all indices. In equation \eqref{eq:equation-spatial-moment-P2}, $\bM_m$ for $m <0$ is understood to be zero.

At contact, $r_2 = R_c$, the no-flux boundary condition is satisfied:
\begin{align}
    &\bn_2\cdot\left(\bU_r^0 \bM_m - D_r^T \nabla_2^T\bM_m + \frac{k}{\zeta_1}\bM_{m+1}\right)\nonumber \\
    &- m D_1^T\left[\bn_2 \bM_{m-1} \right]^\mathrm{sym}=\bm{0}.
\end{align}
The far-field condition for the spatial moment of rank $m$ is
\begin{equation}
    \label{eq:spatial-moment-P2-far-field}
    \bM_m \to \frac{n_b}{\Omega_b} \bPhi_m(\bq_1, t) \quad \mathrm{as}\quad r_2 \to \infty,
\end{equation}
where
\begin{equation}
    \label{eq:P1-spatial-moment-def}
    \bPhi_m(\bq_1, t) = \int  \underbrace{\br_1\cdot\cdot\cdot\br_1}_{m} P_1(\br_1, \bq_1, t) d \br_1
\end{equation}
is the rank $m$ spatial moment of the single particle probability $P_1$ of the probe. Discussion of the single particle behavior and the method to obtain $\bPhi_m$ is deferred to section \ref{sec:single-probe}.

From \eqref{eq:md-pair} and \eqref{eq:msd-pair}, to obtain the mean and mean-squared displacements, one only needs to calculate the zeroth and first spatial moments, respectively. On the other hand, the definitions of mean and mean-squared displacements allow us to write $\avg{\br_1} = \int \bM_1/(N-1) d\bq_1d\Gamma_2$ and $\avg{\br_1\br_1} = \int \bM_2/(N-1) d\bq_1d\Gamma_2$. Because in obtaining $\avg{\br_1\br_1}$ only the integral of $\bM_2$ is required, it's not necessary to first calculate the distribution of $\bM_2$ explicitly before carrying out the integration. Instead, one can show that integrating equation \eqref{eq:equation-spatial-moment-P2} for $m=2$ leads to the same equation as \eqref{eq:msd-pair}. Due to the presence of the harmonic trap force, the equation for $\bM_m$ is coupled to $\bM_{m+1}$. To truncate this infinite set of equations and obtain a finite set of closed equations, a closure model may be used.

To see the structure of the spatial moments more clearly, we write out the first few moment equations explicitly using \eqref{eq:equation-spatial-moment-P2}. The zeroth moment, $M_0 = \int \rho_2 d \br_1$, satisfies the equation
\begin{align}
    \label{eq:pair-M0-equation}
    \frac{\partial M_0}{\partial t} &+ \nabla_2^T\cdot\left[\bU_r^0 M_0
     - D_r^T\nabla_2^T M_0 + \frac{k}{\zeta_1}\bM_1 \right] \nonumber\\
     &- \sum_{\alpha=1}^2 D_\alpha^R \nabla_\alpha^R\cdot\nabla_\alpha^R M_0=0,
\end{align}
and the normalization $\int M_0d\bq_1d\Gamma_2 = N-1$. In addition to being advected by the relative velocity $\bU_r^0$ in the physical space of the bath particle, $M_0$ is forced by the trap via the divergence of the first moment. 

The equation governing the evolution of the first spatial moment is
\begin{eqnarray}
    &&\frac{\partial \bM_1}{\partial t} - \left(U_1^0 \bq_1 M_0 -\frac{k}{\zeta_1} \bM_1 + D_1^T \nabla_2^T M_0 \right) +\bUtrap M_0\nonumber \\
    &&+\nabla_2^T\cdot\left[\bU_r^0 \bM_1 - D^T_r\nabla_2^T\bM_1 + \frac{k}{\zeta_1}\bM_2 -D_1^T\bI M_0\right] \nonumber \\
    &&-\sum_{\alpha=1}^2 D_\alpha^R \nabla_\alpha^R\cdot\nabla_\alpha^R \bM_1=\mathbf{0}.
    \label{eq:pair-M1-equation}
\end{eqnarray}
Similarly, the second moment is governed by
\begin{widetext}
    \begin{align}
         \frac{\partial \bM_2}{\partial t} &-2\left[ U_1^0\bq_1 \bM_1 - \frac{k}{\zeta_1}\bM_2 +D_1^T M_0\bI +D_1^T \nabla_2^T\bM_1 - \bUtrap\bM_1 \right]^\mathrm{sym} \nonumber \\
        &  +\nabla_2^T\cdot\left[\bU_r^0 \bM_2 -D_r^T \nabla_2^T \bM_2 + \frac{k}{\zeta_1} \bM_3 \right]   - 2 D_1^T \left[\nabla_2^T \bM_1\right]^\mathrm{sym} 
        -\sum_{\alpha=1}^2 D_\alpha^R \nabla_\alpha^R\cdot\nabla_\alpha^R \bM_2=\mathbf{0}.
        \label{eq:M2equation}
    \end{align}    
\end{widetext}

\subsection{The probe distribution in the absence of bath particles}
\label{sec:single-probe}
The simplest problem in the above formulation is that of a single particle (the probe) interacting with the trap. One can formulate this single-particle problem by neglecting all bath particles or taking the limit $\phi_b = 4\pi b^3n_b/3 \to 0$ in the above $N$-particle formulation. The single-particle probability $P_1(\br_1, \bq_1, t \rvert \bx_0, t)$ of the active probe satisfies
\begin{align}
    \label{eq:single-particle-P1-eq}
    \frac{\partial P_1}{\partial t}&+\nabla_1^T\cdot\left(\frac{1}{\zeta_1}\bF_1^e P_1 - D_1^T\nabla P_1 - \bUtrap P_1 +U_1^0 \bq_1 P_1 \right)\nonumber\\
    & - D_1^R \nabla_1^R\cdot\nabla_1^R P_1 =0,
\end{align}
where the conservation of probability dictates that $\int P_1 d\Gamma_1 =1$ and the harmonic trap force $\bF_1^e = -k \br_1$. We emphasize that in equation \eqref{eq:single-particle-P1-eq} the probe is also considered as an ABP.

The rank $m$ ($m=0,1,...$) spatial moment of $P_1$ defined by  \eqref{eq:P1-spatial-moment-def} satisfies
\begin{align}
    &\frac{\partial \bPhi_m}{\partial t}\nonumber\\
    & - m \left[U_1^0\bq_1\bPhi_{m-1} - \frac{k}{\zeta_1}\bPhi_m +(m-1)D_1^T \bPhi_{m-2}\bI - \bUtrap\bPhi_{m-1}  \right]^\mathrm{sym} \nonumber \\
    &- D_1^R\nabla_1^R\cdot\nabla_1^R \bPhi_m = \bm{0},
\end{align}
where $\bPhi_m$ for $m <0$ is defined to be zero. Different from equation \eqref{eq:equation-spatial-moment-P2} in which the moment $\bM_m$ is coupled to $\bM_{m+1}$, the rank $m$ spatial moment of $P_1$ only depends on lower order moments, which leads to a set of closed equations. The solution to the preceding equation provides the far-field condition for $\bM_m$ as given by equation \eqref{eq:spatial-moment-P2-far-field}.

The zeroth-order spatial moment $\Phi_0$ is the net orientational distribution, which is unaffected by the trap and is governed by the orientational diffusion equation:
\begin{equation}
    \frac{\partial \Phi_0}{\partial t} - D_1^R \nabla_1^R\cdot\nabla_1^R \Phi_0=0,
\end{equation}
where the conservation of $P_1$ gives  $\int \Phi_0 d \bq_1 = 1$. At long times, the solution is simply the uniform distribution, $\Phi_0(\bq_1, t \to \infty) = 1/(4\pi)$ in 3D.


The above formulation also allows us to consider the mean and fluctuation of the probe displacement in the absence of bath particles. Equation \eqref{eq:md-PN} or \eqref{eq:md-pair} in the absence of bath particles reduces to
\begin{equation}
    \label{eq:md-P1}
    \frac{\partial \avg{\br_1}}{\partial t}+\frac{1}{\tau_k} \avg{\br_1}  = -\bUtrap+ U_1^0 \avg{\bq_1},
\end{equation}
where for the single particle $\avg{\br_1} = \int \br_1P_1d\Gamma_1 = \int \bPhi_1 d\bq_1$. Similarly, equation \eqref{eq:msd-PN} or \eqref{eq:msd-pair} for the single particle becomes
\begin{equation}
    \label{eq:msd-P1}
    \frac{1}{2}\frac{\partial \avg{\br_1\br_1}}{\partial t}    +\frac{1}{\tau_k}\bigavg{\br_1\br_1}=D_1^T \bI  +\left[U_1^0 \bigavg{\bq_1\br_1} - \bUtrap\avg{\br_1}  \right]^\mathrm{sym}.
\end{equation}

It can be seen from equations \eqref{eq:md-P1} and \eqref{eq:msd-P1} that in order to calculate the mean and mean-squared displacements, one needs to obtain the net polar order $\avg{\bq_1}$ and the covariance of the position and orientation $\Cov(\bq_1,\br_1)$. The governing equation for  $\Cov(\bq_1,\br_1)$ follows from \eqref{eq:Covariance-q1-r1-pair} and is given by
\begin{equation}
    \label{eq:covariance-q1-r1-single}
    \frac{\partial \Cov(\bq_1, \br_1)}{\partial t} + \frac{1}{\tau}\Cov(\bq_1, \br_1) =U_1^0\Var(\bq_1)
\end{equation}
which depends on the net nematic order $\avg{\bq_1\bq_1}$.

At steady state, it is shown that $\avg{\bq_1} = \bm{0}$ and $\avg{\bq_1\bq_1} = \bI/d$, where $d=2,3$ is the dimensionality of the physical space. This allows us to obtain
\begin{align}
    \label{eq:mean-single-particle}
    \avg{\br_1} &= -\frac{\zeta_1 \bUtrap}{k},\\
    \label{eq:covariance-q1-r1-single-result}
    \Cov(\bq_1,\br_1) &= \frac{U_1^0}{dk/\zeta_1 + d(d-1)D_1^R} \bI,
\end{align}
\begin{align}
    \label{eq:covariance-r1-r1-single}
    \avg{\br_1\br_1} = &\frac{\zeta_1^2}{k^2} \bUtrap\bUtrap + \frac{\zeta_1D_1^T}{k}\bI \nonumber\\
    &+ \frac{\zeta_1 D_1^\mathrm{swim}}{k} \frac{1}{1 + \frac{k \tau_1^R}{\zeta_1} \frac{1}{d-1}}\bI,
\end{align}
where $D_1^\mathrm{swim} = \left(U_1^0\right)^2\tau_1^R/[d(d-1)]$ is the swim diffusivity of a freely swimming ABP. The average position of the ABP relative to the trap is given by the balance between the average trap force $k\avg{\br_1}$ and the viscous drag $\zeta_1\bUtrap$. If the trap is strong, $k\to \infty$, the ABP is tightly confined and pushing against the trap `boundary', which has been observed in experiments \citep{takatori2016acoustic}. On the other hand, for $k \to 0$, the average position of the ABP becomes unbounded. Solving the steady state first and then taking the limit $k\to 0$ in \eqref{eq:mean-single-particle} is singular because in the absence of the trap ($k=0$) the average position is unbounded and at long times the particle motion is diffusive. For $k\equiv 0$, we are simply measuring the motion of an ABP in a frame of reference moving with velocity $\bUtrap$ relative to the laboratory frame, which gives $d \avg{\br_1}/dt = -\bUtrap$.

\citet{takatori2016acoustic} studied the transient and long-time dynamics of self-propelled Janus particles in a fixed acoustic trap. They showed that the experimentally measured density distribution of Janus particles follow closely the theoretical predictions using a harmonic trap. Equation \eqref{eq:covariance-r1-r1-single} in the absence of $\bUtrap$ agrees with that obtained in \citet{takatori2016acoustic}.

The fluctuation relation is given by
\begin{equation}
    \label{eq:Force-displacement-fluctuation}
    \bigavg{\left(\Delta\bF_1^e \right)^2}^{1/2}\bigavg{\left(\Delta \br_1\right)^2}^{1/2} = d \left[\kt + \frac{\ksts}{1+ \tau_1^R/[(d-1)\tau_k]}\right],
\end{equation}
where the thermal energy $\kt = \zeta_1D_1^T$ and analogously an \emph{active} energy scale $\ksts$ has been defined such that $\ksts = \zeta_1D_1^\mathrm{swim}$ \citep{takatoriprl14}. In equation \eqref{eq:Force-displacement-fluctuation}, the fluctuation consists of the thermal (passive) energy $d\kt$  and an active energy contribution. This active energy is different from $\ksts$ due to the presence of the harmonic trap, which introduces an orientational decorrelation timescale $\tau_k$  in addition to the reorientation time $\tau_1^R$ of the ABP. For a weak trap, $\tau_1^R/\tau_k \ll 1$, the decorrelation occurs on the timescale of $\tau_1^R$, and the active contribution scales as $\zeta_1\left(U_1^0\right)^2\tau_1^R$. As a result, the fluctuation $\bigavg{\left(\Delta\bF_1^e \right)^2}^{1/2}\bigavg{\left(\Delta \br_1\right)^2}^{1/2} \to d(\kt+\ksts)$ as $\tau_1^R/\tau_k \to 0$. This is often referred to as Rule \#1 of active matter---when all length scales are large compared to the run length $\ell_1$, one can replace $\kt$ with $\kt + \ksts$. As another example, consider the sedimentation of active colloids under gravity. At steady state, the number density follows  Boltzmann distribution but with $\kt + \ksts$ in place of $\kt$ \citep{Palacci2010}. 

When $\tau_1^R/\tau_k \gg 1$, the relevant timescale is $\tau_k$, and the active contribution scales as $\zeta_1 \left(U_1^0\right)^2\tau_k$. In this limit, the ABP is pushing against  the edge of the potential well and the fluctuation comes from passive Brownian motion alone, $\bigavg{\left(\Delta\bF_1^e \right)^2}^{1/2}\bigavg{\left(\Delta \br_1\right)^2}^{1/2} \to d \kt$ as $\tau_1^R/\tau_k \to \infty$.

Regardless of the trap strength, the product of the square root of the fluctuations in the force and the position is always bounded. For a strong trap, the position fluctuation vanishes, $\bigavg{\Delta\br_1\Delta\br_1} = O(1/k) \to 0$, but the force fluctuation blows up linearly since $\bigavg{\Delta\bF_1^e\Delta\bF_1^e} = O(k) \to \infty$ as $k \to \infty$. Conversely, the position fluctuation grows unboundedly while the force fluctuation vanishes as $k \to 0$.

In the weak trap limit, equation \eqref{eq:Force-displacement-fluctuation} can be equivalently written as
\begin{equation}
    \label{eq:fluctuation-diffusion-equivalence}
    \frac{k}{\zeta_1}\bigavg{\Delta\br_1\Delta\br_1}= \frac{\bigavg{\Delta\br_1\Delta\br_1}}{\tau_k} \to \bD_1^\mathrm{eff}\quad\mathrm{as}\quad \frac{\tau_1^R}{\tau_k}\to 0,
\end{equation}
where $\bD_1^\mathrm{eff}= D_1^T\bI +D_1^\mathrm{swim}\bI$ is the long-time effective diffusivity of the ABP in the absence of the trap (see appendix \ref{sec:appendix-transient-probe} for the asymptotic analysis). This relation implies the equivalence of the position fluctuation divided by $\tau_k$ in the limit of vanishing harmonic trapping force and the effective diffusion of a free ABP. In other words, one could calculate the position fluctuation in a trap and then take the limit of $\bigavg{\Delta\br_1\Delta\br_1}/\tau_k$ as $k\to 0$  to obtain the long-time diffusivity that the particle would have in the absence of the trap, or vice versa. Because the trap is weak, the ABP is able to explore space via both thermal fluctuation and its undisturbed active run-and-reorientation, both processes contribute to the position fluctuation. In the presence of bath particles, this equivalence still holds in which $\bD_1^\mathrm{eff}$ is the diffusivity of the probe affected by collisions with bath particles (i.e., tracer diffusion).

\subsection{A weak trap}
For a weak trap, $\epsilon= \tau_1^R/\tau_k = k \tau_1^R/\zeta_1 \ll 1$, the probe is allowed to explore and reorient freely before reaching the ``boundary'' of the potential well. The viscoelastic timescale $\tau_k$ is well separated from the reorientation timescale $\tau_1^R$. In the intermediate timescale characterized by $t/\tau_1^R \gg 1$ and $t/\tau_k \ll 1$,  the probe has explored the suspension but has not reached the boundary of the potential; we expect a diffusive behavior of the probe. At times much longer than the viscoelastic timescale ($t/\tau_k \gg 1$), the variance of the probe position becomes bounded due to the trapping force. Therefore, the motion of the probe exhibits a transition from diffusive to bounded behavior.

The separation of the two timescales allows us to consider a multiple-scale analysis. By defining the fast variable $t_1 = t$ and the slow variable $t_2 = \epsilon t$, we have $\partial/\partial t = \partial/\partial t_1 + \epsilon \partial/\partial t_2$. Regular perturbation expansions of the pair probability distribution and its spatial moments in terms of $\epsilon$ are written as
\begin{eqnarray}
    \rho_2 &=&\rho_2^{(0)}+\epsilon \rho_2^{(1)}+\cdot\cdot\cdot,\\
    \bM_m &=& \bM_m^{(0)} + \epsilon \bM_m^{(0)}+\cdot\cdot\cdot,
\end{eqnarray}
where $\bM_m^{(k)}$ is the rank $m$ spatial moment of $\rho_2^{(k)}$.

At $O(1)$, the zeroth moment satisfies
\begin{align}
    \label{eq:weak-trap-M00}
    &\frac{\partial M_0^{(0)}}{\partial t_1} + \nabla_2^T\cdot\left(\bU_r^0 M_0^{(0)}
    - D_r^T\nabla_2^T M_0^{(0)} \right) \nonumber\\
    &  - \sum_{\alpha=1}^2 D_\alpha^R \nabla_\alpha^R\cdot\nabla_\alpha^R M_0^{(0)}=0,\\
    &\bn_2\cdot\left(\bU_r^0 M_0^{(0)} - D_r^T\nabla_2^T M_0^{(0)} \right) = 0,\quad \br_2 \in S_c.
\end{align}
Similarly, the first moment at this order is given by
\begin{align}
    \label{eq:weak-trap-M10}
    &\frac{\partial \bM_1^{(0)}}{\partial t_1} - \left(U_1^0 \bq_1 M_0^{(0)}  + D_1^T \nabla_2^T M_0^{(0)} \right) +\bUtrap M_0^{(0)}\nonumber \\
    &+\nabla_2^T\cdot\left(\bU_r^0 \bM_1^{(0)} - D^T_r\nabla_2^T\bM_1^{(0)}  -D_1^T\bI M_0^{(0)}\right)\nonumber \\
    &-\sum_{\alpha=1}^2 D_\alpha^R \nabla_\alpha^R\cdot\nabla_\alpha^R \bM_1^{(0)}=\mathbf{0},\\
    & \bn_2\cdot\left(\bU_r^0 \bM_1^{(0)} - D^T_r\nabla_2^T\bM_1^{(0)}  -D_1^T\bI M_0^{(0)}\right)=0,\quad \br_2 \in S_c.
\end{align}

Expanding the covariances similarly, e.g., 
\begin{equation}
    \Var(\br_1) = \Var^{(0)}(\br_1) + \epsilon \Var^{(1)}(\br_1)+\cdot\cdot\cdot,
\end{equation}
we obtain at $O(1)$
\begin{widetext}
    \begin{align}
        \label{eq:weak-trap-covariance-r1-r1}
        \frac{1}{2}\frac{\partial \Var^{(0)}(\br_1)}{\partial t_1}= D_1^T\bI +\Big[ U_1^0\Cov^{(0)}(\bq_1, \br_1)+ D_1^T\int \nabla_2^T\rho_2^{(0)}\Delta \br_1 d\Gamma^2   \Big]^\mathrm{sym},
    \end{align}
    \begin{align}
        \label{eq:weak-trap-covariance-q1-r1}
        \frac{\partial \Cov^{(0)}(\bq_1, \br_1)}{\partial t_1}  +\frac{d-1}{\tau_1^R} \Cov^{(0)}(\bq_1, \br_1) =U_1^0\Var(\bq_1) + D_1^T \int \Delta\bq_1\nabla_2^T\rho_2^{(0)} d\Gamma^2.
    \end{align}    
\end{widetext}
Note that $\Var(\bq_1)$ is not affected by the presence of the trap (see appendix \ref{sec:appendix-pair-covariance}) and therefore only has the $O(1)$ term in the small-$\epsilon$ expansion.

Equations \eqref{eq:weak-trap-M00}--\eqref{eq:weak-trap-covariance-q1-r1} govern the dynamics of a probe in a bath of active particles in the absence of the trapping force (The presence of $\bUtrap$ in \eqref{eq:weak-trap-M10} is only due to the fact that we are in a frame of reference moving with $\bUtrap$ relative to the laboratory frame). This problem is the so-called tracer---an active one---diffusion in an active Brownian suspension. Even in the absence of the trap, the correlation between $\bq_1$ and $\br_1$ has a steady-state (time-independent) solution due to the presence of the decorrelation time $\tau_1^R$ in equation \eqref{eq:weak-trap-covariance-q1-r1}. Dropping the time derivative in \eqref{eq:weak-trap-covariance-q1-r1} at steady state, we obtain
\begin{align}
    \Cov^{(0)}(\bq_1, \br_1) = &\frac{\ell_1}{d(d-1)}\bI \nonumber \\
    &+ \frac{\tau_1^R}{d-1}D_1^T \int \Delta\bq_1\nabla_2^T\rho_2^{(0)} d\Gamma^2,
\end{align}
where it is understood that the steady-state distribution of $\rho_2^{(0)}$ is used, and $\ell_1=U_1^0\tau_1^R$ is the run length of the active probe. Therefore, equation \eqref{eq:weak-trap-covariance-r1-r1} is written as
\begin{align}
    \label{eq:weak-trap-Deff}
    \bD_1^\mathrm{eff}  = & \left( D_1^T +D_1^\mathrm{swim}\right)\bI \nonumber\\
    &+ D_1^T\left[\int \left( \frac{\ell_1}{d-1} \Delta\bq_1 +\Delta \br_1 \right)\nabla_2^T \rho_2^{(0)}d\Gamma^2 \right]^\mathrm{sym},
\end{align}
where $\bD_1^\mathrm{eff}=\partial \Cov^{(0)}(\br_1,\br_1)/(2\partial t_1)$ is the long-time diffusivity of the probe in the absence of the trapping force. As expected, one could obtain the same result by setting $\bF_1^e = \bm{0}$ from the outset (see section \ref{sec:tracer-diffusion}). This is done in \citet{Burkholder2017} but with the free tracer particle being passive.

So long as the trapping force is not identically zero, the probe will eventually reach the boundary of the trap.  This confinement happens at very large distances from the trap (or at long times if the probe is started near the trap center).

\subsection{A strong trap}
For a strong trap, the viscoelastic time scale $\tau_k = \zeta_1/k$ is much smaller than other timescales (e.g., the reorientation time ) of the problem. Due to the strong trapping force, both the mean and the variance of the probe have a steady-state solution that is time independent.

The position fluctuation, governed by equation \eqref{eq:var-r1-q1r1-text}, becomes at steady state
\begin{align}
    \label{eq:covariance-r1-r1-strongtrap}
    \frac{k}{\zeta_1}\Var( \br_1) = &D_1^T\bI + \Big[ U_1^0\Cov(\bq_1, \br_1)\nonumber\\
    &+D_1^T\int \Delta \br_1 \nabla_2^T \rho_2d\Gamma^2  \Big]^\mathrm{sym}.
\end{align}
Similarly, $\Cov(\bq_1, \br_1)$ defined in \eqref{eq:Covariance-q1-r1-pair} is given by
\begin{equation}
    \frac{k}{\zeta_1} \Cov(\bq_1, \br_1) = \frac{1}{d}U_1^0 \bI + D_1^T\int \Delta \bq_1 \nabla_2^T \rho_2d\Gamma^2.
\end{equation}
Because the last term in the preceding equation is finite as $k\to \infty$, $\Cov(\bq_1, \br_1)$ is small and on the order of $1/k$.

On the other hand, for a strong trap, the relative deviation of the probe position from the average position is small, $\Delta|\br_1|/\avg{|\br_1|} \ll 1$, which leads to
\begin{align}
    \label{eq:strong-trap-small-deviation}
    \bM_1(\br_2, \bq_2, t) &= \int \br_1 \rho_2 d\br_1 = \bigavg{\br_1} M_0 + \int \Delta \br_1 \rho_2 d\br_1 \nonumber \\
    &=  \bigavg{\br_1} M_0 + O(\Delta |\br_1|/\avg{|\br_1|} ),
\end{align}
where the decomposition $\br_1 = \avg{\br_1}+\Delta \br_1$ is used. Using the first line of \eqref{eq:strong-trap-small-deviation}, we have 
\begin{equation}
\int \nabla_2^T \rho_2 \Delta \br_1 d\Gamma^2 =\int \nabla_2^T\big[\bM_1- \avg{\br_1}M_0\big]d \bq_1 d\Gamma_2,
\end{equation}
which is negligible due to the second line of  \eqref{eq:strong-trap-small-deviation}. Taken together, we conclude that the last two terms on the rhs of \eqref{eq:covariance-r1-r1-strongtrap} are subdominant. To leading-order, the fluctuation in the strong-trap limit is given by
\begin{equation}
    k \Var(\br_1) = \zeta_1 D_1^T\bI = \kt \bI,
\end{equation}
regardless of the presence of the bath particles. Therefore, in this limit we have 
\begin{align}
    \bigavg{\left(\Delta\bF_1^e \right)^2}^{1/2}\bigavg{\left(\Delta \br_1\right)^2}^{1/2} = d \kt.
\end{align}

\section{Constant-force and constant-velocity microrheology}
\label{sec:CF-and-CV}
In this section, we show that the trapped-particle microrheology problem can be reduced to either the CV or CF problem when appropriate limits are taken.

\subsection{Constant-force microrheology}
To recover the constant-force microrheology problem, instead of a harmonic trapping force, we apply a constant force to the probe particle, $\bF_1^e = const$,  and set the trap velocity $\bUtrap=\bm{0}$. In this mode of operation, the main quantity of interest is the average velocity $\avg{\bU_1}$ of the probe in response to the constant external driving force. By definition,  $\avg{\bU_1} = \partial \avg{\br_1}/\partial t$, which can be obtained by considering the rhs of equation \eqref{eq:md-PN}.

Because the trap is absent, the position $\br_1$ defines an arbitrary origin in the laboratory frame of reference and the system is statistically homogeneous \citep{SquiresBrady2005}. As a result, the conditional probability $P_{N-1/1}$ defined by
\begin{equation}
    \label{eq:PDF-conditioned-on-P1}
    P_N = P_{N-1/1}(\br^{N-1}, \bq^{N-1}, t \rvert \br_1, \bq_1, t)P_{1}( \br_1, \bq_1, t)
\end{equation}
is not a function of $\br_1$. (Note that in general $P_{N-1/1}$ can be a function of $\bq_1$.) The third term on the rhs of equation \eqref{eq:md-PN} becomes
\begin{align}
    - \bigavg{ \bD_{11}^{UF}\cdot\nabla_1^T\ln P_N} &= - \int d \Gamma^{N-1} P_{N-1/1}\bD_{11}^{UF} \cdot \int d \Gamma_1 \nabla_1^T P_1 \nonumber\\
    &= \bm{0},
\end{align}
where we have used the divergence theorem and the fact that $P_1$ vanishes at infinity.

Further manipulations allow us to write equation \eqref{eq:md-PN} as
\begin{align}
    \label{eq:avg-U1-CF}
    \avg{\bU_1} =& U_1^0 \avg{\bq_1} + \bigavg{\bM_{11}^{UF}}\cdot\bF_1^e \nonumber \\
    &- \sum_{\beta=1}^N\int d\Gamma_1 P_1\int d\Gamma^{N-1}\left(\bD_{1\beta}^{UF} - \bD_{1 1}^{UF}\right)\cdot \nabla_\beta^T  P_{N-1/1} .
\end{align}
If all $N$ particles (including both the probe and the bath particles) are passive, equation \eqref{eq:avg-U1-CF} upon integration over $\Gamma_1$  reduces to the average velocity relation originally obtained by \citet{SquiresBrady2005} (equation A4) for passive colloids.

Neglecting hydrodynamic interactions in the dilute limit, the average velocity becomes
\begin{equation}
    \label{eq:avg-U1-CF-pair}
    \avg{\bU_1} =  \frac{1}{\zeta_1}\bF_1^e + U_1^0 \avg{\bq_1}+ D_1^T \int \nabla_2^T \rho_2d\Gamma^2.
\end{equation}
Recalling that $M_0 = \int \rho_2 d\br_1$, the last term in \eqref{eq:avg-U1-CF-pair}  can be calculated so long as $M_0$ can be obtained. We note that, at long times, $\avg{\bq_1}=\bm{0}$. If the probe is under the influence of external orienting fields, the net polar order $\avg{\bq_1}$ becomes non-zero \citep{Takatori2014sm}.

In the CF mode of microrheology,  the equation governing the spatial moment $\bM_m$ is similar  to \eqref{eq:equation-spatial-moment-P2} and can be shown to be
\begin{widetext}
    \begin{eqnarray}
        &&\frac{\partial \bM_m}{\partial t} - m \left[U_1^0\bq_1 \bM_{m-1} + \frac{1}{\zeta_1}\bF_1^e\bM_{m-1} +(m-1) D_1^T \bM_{m-2}\bI +D_1^T \nabla_2^T\bM_{m-1} \right]^\mathrm{sym}\nonumber \\
        && +\nabla_2^T\cdot\left(\bU_r^0 \bM_m - D_r^T \nabla_2^T\bM_m - \frac{1}{\zeta_1}\bF_1^e\bM_{m}\right) - m D_1^T \left[\nabla_2^T \bM_{m-1} \right]^\mathrm{sym} - \sum_{\alpha=1}^2 D_\alpha^R\nabla_\alpha^R\cdot\nabla_\alpha^R \bM_m=\bm{0}.
        \label{eq:equation-spatial-moment-CF}
    \end{eqnarray}
\end{widetext}
Here, because the external force is constant, the moment equation at rank $m$ only depends on moments of lower ranks and the system up to any rank is a closed set of equations.

At contact, $r_2 = R_c$, we have the no-flux boundary condition:
\begin{align}
    &\bn_2\cdot\left(\bU_r^0 \bM_m - D_r^T \nabla_2^T\bM_m - \frac{1}{\zeta_1}\bF_1^e\bM_{m}\right) \nonumber\\
    &- m D_1^T\left[\bn_2 \bM_{m-1} \right]^\mathrm{sym}=\bm{0}.
\end{align}
The far-field condition as $r_2\to \infty$ is unchanged and given by equation \eqref{eq:spatial-moment-P2-far-field}, where $\bPhi_m$ for constant force satisfies
\begin{align}
    \label{eq:Phi-m-CF}
    &\frac{\partial \bPhi_m}{\partial t} - m \left[U_1^0\bq_1\bPhi_{m-1} + \frac{\bF_1^e}{\zeta_1}\bPhi_{m-1} +(m-1)D_1^T \bPhi_{m-2}\bI  \right]^\mathrm{sym} \nonumber\\
    &- D_1^R\nabla_1^R\cdot\nabla_1^R \bPhi_m = \bm{0}.
\end{align}
To find the average velocity given in equation \eqref{eq:avg-U1-CF-pair}, one needs to consider equations \eqref{eq:equation-spatial-moment-CF} and \eqref{eq:Phi-m-CF} for $m=0$. 

We note that in the above general formulation, both the probe particle and the bath particle are ABPs. By setting $U_1^0, U_2^0=0$ and integrating out the orientational degrees of freedom of the probe and the bath particle, we obtain the  CF microrheology problem of a passive Brownian probe in a passive Brownian suspension, which has been considered by \citet{SquiresBrady2005}. On the other hand, the CF microrheology of a passive Brownian probe in an active Brownian suspension ($U_1^0=0, U_2^0\neq 0$) is studied by \citet{Burkholder2019JCP}.

Taking $m=0$ in equation \eqref{eq:equation-spatial-moment-CF} in the absence of the external force ($\bF_1^e=\bm{0}$), we obtain
\begin{align}
    \label{eq:M0-no-force}
    &\frac{\partial M_0}{\partial t}  +\nabla_2^T\cdot\left(\bU_r^0 M_0 - D_r^T \nabla_2^T M_0 \right) \nonumber\\
    &- \sum_{\alpha=1}^2 D_\alpha^R\nabla_\alpha^R\cdot\nabla_\alpha^R M_0=0.
\end{align}
Treating the probe as one of the suspension particles, this zeroth spatial moment is the pair-correlation function of an active Brownian suspension (subject to proper normalization) in the dilute limit by neglecting all higher order correlations. Equation \eqref{eq:M0-no-force} governing the pair-correlation at steady state in 2D has been studied \citep{Poncet2021, Dhont2021}.

\subsection{Force-induced tracer diffusion}
\label{sec:tracer-diffusion}
In the constant-force mode of microrheology, it is also of importance to consider the force-induced diffusion of the probe  particle. In this context, the probe is often referred to as the tracer, i.e., force-induced tracer diffusion. If no external force is applied,  $\bF_1^e = \bm{0}$, the problem is simply called tracer diffusion. The long-time diffusivity of the tracer in the presence of bath particles can be written as
\begin{align}
    \bD_1^\mathrm{eff} &=  \lim_{t\to \infty}\frac{1}{2} \frac{d}{dt} \Var(\br_1) \nonumber\\
    &=\lim_{t\to \infty}\frac{1}{2} \left[\frac{\partial }{\partial t}\avg{\br_1\br_1} - \avg{\bU_1}\avg{\br_1} - \avg{\br_1}\avg{\bU_1} \right],
\end{align}
where the covariance tensor of $\br_1$ is governed by
\begin{align}
    \label{eq:covariance-r1-r1-CF}
    \frac{1}{2}\frac{d }{d t}\Var(\br_1) =& D_1^T\bI +U_1^0\big[\Cov(\bq_1, \br_1) \big]^\mathrm{sym}\nonumber\\
    &+D_1^T\Big[\int \Delta\br_1\nabla_2^T \rho_2d\Gamma^2  \Big]^\mathrm{sym},
\end{align}
and the covariance of $\bq_1$ and $\br_1$ satisfies
\begin{align}
    \label{eq:covariance-q1-r1-CF}
    \frac{d}{d t} \Cov(\bq_1, \br_1)  +\frac{d-1}{\tau_1^R} \Cov(\bq_1, \br_1) =&U_1^0\Var(\bq_1) \nonumber \\
    &+ D_1^T\int \Delta\bq_1\nabla_2^T \rho_2d\Gamma^2.
\end{align}

At long times,  we then obtain the diffusivity as
\begin{align}
    \label{eq:CF-Deff}
    \bD_1^\mathrm{eff} =& \left( D_1^T +D_1^\mathrm{swim}\right)\bI \nonumber\\
    &+ D_1^T\left[\int \left( \frac{\ell_1}{d-1} \Delta\bq_1 +\Delta \br_1 \right)\nabla_2^T \rho_2 d\Gamma^2 \right]^\mathrm{sym}.
\end{align}
In \eqref{eq:CF-Deff}, the first bracketed term on the rhs is the diffusivity of a single ABP in free space and the remaining terms are the additional contributions due to the excluded-volume interaction with the bath particles. As alluded to earlier, equation \eqref{eq:CF-Deff} is identical to equation \eqref{eq:weak-trap-Deff},  which is obtained in the weak-trap limit. We note that in  \eqref{eq:CF-Deff} there is a constant external force while the diffusivity obtained in \eqref{eq:weak-trap-Deff} is for a free tracer, i.e., force-induced versus free tracer diffusion. It is clear that if the force is absent the diffusivities obtained from \eqref{eq:weak-trap-Deff} and \eqref{eq:CF-Deff}  are identical.

Using the divergence theorem, we can relate the integrals on the rhs of  \eqref{eq:CF-Deff} to the zeroth and first spatial moments,
\begin{align}
    \int \Delta\bq_1\nabla_2^T \rho_2d\Gamma^2 =\int \Delta\bq_1 d\bq_1d\bq_2\oint_{S_c} \bn_2M_0 dS_2,
\end{align}
\begin{align}
    \int \Delta\br_1\nabla_2^T \rho_2d\Gamma^2 =\int \Delta\bq_1 d\bq_1d\bq_2\oint_{S_c}\big( \bM_1- \avg{\br_1}M_0\big)\bn_2 dS_2,
\end{align}
where $\avg{\br_1} = \int \bM_1/(N-1) d\bq_1d\Gamma_2$. Therefore, one only needs to solve for $M_0$  and $\bM_1$ in equation \eqref{eq:equation-spatial-moment-CF} in order to calculate the diffusivity.

The above formulation for the forced-induced diffusion of an active tracer in an active suspension is a direct extension of the generalized Taylor dispersion theory (GTDT). In particular, we have used the statistical moment method of  \citet{frankel_brenner_1989}. An equivalent approach is to derive the mean velocity and the diffusivity by first transforming the unbounded coordinate $\br_1$ into the Fourier space and consider a small wave-number expansion \citep{zia_brady_2010,Burkholder2017,Burkholder2019JCP}.

By setting $U_1^0, U_2^0 =0$ and integrating over the orientational degrees of freedom of both the probe and the bath particles,  we recover the equations governing the force-induced diffusion of a passive probe in a passive suspension \citep{zia_brady_2010}. To recover the problem of a passive free tracer in an active suspension studied by  \citet{Burkholder2017}, one can set $\bF_1^e = \bm{0}$,  $U_1^0=0$ and integrate over the orientational degrees of freedom of the probe.

\subsection{Constant-velocity microrheology}
To obtain the equations for the CV microrheology problem, we first consider the probe to have deterministic dynamics with $U_1^0 =0$, $D_1^T=0$ and $D_1^R=0$. Equation \eqref{eq:md-pair} at steady-state then leads to $k\avg{\br_1}/\zeta_1 = -\bUtrap$. Furthermore, we consider the limit of a strong trap in which case the probe tightly follows the trap velocity. In this limit, the probe velocity is the trap velocity to leading-order and we then achieve a CV probe.

To see this, we first decompose the position of the probe via $\br_1 = \avg{\br_1}+\Delta \br_1$. In the strong trap limit, the deviation of the probe from the mean position is small, $\Delta |\br_1|/\avg{|\br_1|} \ll 1$.

To leading-order, \eqref{eq:strong-trap-small-deviation} allows us to obtain the first spatial moment as $k\bM_1/\zeta_1 = -\bUtrap M_0$ (this relation can also be viewed as a closure for the spatial moments). Substitution of this relation into \eqref{eq:pair-M0-equation} leads to
\begin{align}
    \label{eq:M0-CV}
    \frac{\partial M_0}{\partial t} &+ \nabla_2^T\cdot\left(U_2^0\bq_2 M_0
     - D_1^T\nabla_2^T M_0 -\bUtrap M_0 \right)\nonumber\\
     &  -  D_2^R \nabla_2^R\cdot\nabla_2^R M_0=0.
\end{align}
Similarly, the no-flux condition at contact ($\br_2 \in S_c$) is
\begin{equation}
    \label{eq:M0-CV-bc}
    \bn_2\cdot\left(U_2^0\bq_2  M_0 - D_1^T\nabla_2^T M_0 -\bUtrap M_0 \right) = 0.
\end{equation}
Equation \eqref{eq:M0-CV} describes the distribution of the bath particle measured in a frame of reference that is co-moving with $\bUtrap$. Realizing that the probe velocity is the same as the `trap', $\bU^\mathrm{probe} = \bUtrap$, this is the CV microrheology of an active Brownian suspension. We note that in \eqref{eq:M0-CV} [cf. \eqref{eq:pair-M0-equation}] the relative velocity is $U_2^0\bq_2 - \bUtrap$ and the relative diffusivity is $D_1^T$ because the probe has prescribed kinematics.

The CV microrheology of an active Brownian suspension governed by \eqref{eq:M0-CV} and \eqref{eq:M0-CV-bc} has been studied by \citet{Burkholder2020} and \citet{peng2022forced}. To recover the CV microrheology of a passive Brownian suspension considered by \citet{SquiresBrady2005}, one only needs to set $U_2^0=0$ and integrate over the orientational degrees of freedom of the bath ABP.

\section{Conclusions}
\label{sec:conclusion}
In this paper we have considered the trapped-particle microrheology of an active colloidal suspension consisting of active Brownian spheres. In the classical models of colloidal microrheology, the applied external force or the probe velocity are fixed and not subject to random fluctuations. This constraint of either CF or CV allows a model simpler than that discussed in the present paper. For the purpose of quantifying the micro-viscous response of suspensions, the CF or CV models are often sufficient. The challenge arises if one wishes to consider the fluctuations of the probe as a result of its interactions with the bath particles and the solvent. More specifically, we have demonstrated that in order to provide a meaningful quantification of the fluctuations in the probe position, one must allow both the position of and the external force on the probe to fluctuate. To achieve this, we developed a generalized microrheology model in which the probe is driven by a translating harmonic trap. We explicitly formulated the equations governing the dynamics of the probe-bath pair in the dilute limit and showed that both the mean position and the fluctuation of the probe position can be given in terms of the joint probability distribution.

In the weak-trap limit, we showed that at an intermediate time the probe exhibits a diffusive behavior in which the diffusivity is the effective diffusivity of a free tracer immersed in the suspension. At this timescale, the probe has explored the suspension but hasn't reached the boundary of the trap. In other words, it is equivalent to the free-tracer diffusion problem. For a strong trap, the fluctuations from the activity of the bath particles or from the collisions between the probe and the bath particles are suppressed due to the strong confinement of the trap. In this limit, the fluctuation of the probe originates from the thermal energy alone regardless of the presence (or activity) of the bath particles.

To conclude, we note that the derived Smoluchowski equation---even at the pair level---has a high dimensionality, which presents a challenge for the computation of the probability density. To circumvent this, one can start from a micromechanical perspective using the Langevin equations and consider a dynamic simulation of the suspension and the probe \citep{Foss2000,Carpen2005,Brady1988}. In a dynamic simulation, the discrete trajectory of the probe is recorded and the calculation of its mean and fluctuation is straightforward.

\begin{acknowledgments}
This work is supported by the National Science Foundation under
Grant No. CBET 1803662.
\end{acknowledgments}

\section*{Data Availability Statement}
The data that support the findings of
this study are available within the
article.

\appendix

\section{Derivation of the pair problem}
\label{sec:pair-derivation}
We integrate equation \eqref{eq:smoluchowski-movingtrap-r-coordinate} over the relative positions and the orientations of the bath particles labeled from $3$ to $N$ to obtain
\begin{align}
    \frac{\partial P_2}{\partial t} &+ \nabla_1^T\cdot\int \left(\bj_1^T - \bUtrap P_N \right)d\Gamma^{N-2} \nonumber\\
    &+ \nabla_2^T\cdot\int\left(\bj_2^T-\bj_1^T\right)d\Gamma^{N-2} \nonumber \\
    &+ \sum_{\alpha=1}^2 \nabla_\alpha^R\cdot\int \bj_\alpha^R d\Gamma^{N-2}=0,
    \label{eq:integrate-3toN}
\end{align}
where $d\Gamma^{N-2}$ is a shorthand for $\prod_{\beta=3}^N d\Gamma_\beta$ and $P_2 = \int P_N d\Gamma^{N-2}$. In deriving the preceding equation, the divergence theorem and the no-flux condition are used to eliminate the terms $\int \nabla_\beta^T\cdot\left(\bj_\beta^T-\bj_1^T\right) d\Gamma^{N-2}$ for $\beta = 3,...,N$. In addition,  the relation $\int \nabla_\alpha^R\cdot\bj_\alpha^R d \bq_\alpha = 0$ is used.

To proceed further, we define the conditional probability of finding the remaining $N-2$ particles, $P_{(N-2)/2}$, given the configuration of the probe and the \emph{first} bath particle:
\begin{widetext}
    \begin{equation}
        P_N\left(\br^N, \bq^N, t\right) = P_{(N-2)/2}\left(\br^{N-2}, \bq^{N-2},t\big\rvert \br_2, \bq_2, \br_1, \bq_1,t \right)P_2(\br_2, \bq_2, \br_1,\bq_1, t).
    \end{equation}   
Notice that the conditional probability is conserved, $\int P_{N-2/2}d\Gamma^{N-2} =1$. In equation \eqref{eq:integrate-3toN}, for $\alpha=1$ or $2$, we have
    \begin{align}
        \label{eq:jT-N-2}
        \int \bj_\alpha^T d \Gamma^{N-2}= & U_\alpha^0 \bq_\alpha P_2 + \bigavg{\bM_{\alpha1}^{UF}}_{(N-2)/2}\cdot\bF_1^e P_2 - \bigavg{\bD_{\alpha2}^{UF}-\bD_{\alpha1}^{UF} }_{(N-2)/2}\cdot\nabla_2^T P_2\nonumber \\
        & - \sum_{\beta=2}^N\Bigavg{\left(\bD_{\alpha\beta}^{UF}-\bD_{\alpha1}^{UF}\right)\cdot\nabla_\beta^T\ln P_{(N-2)/2} }_{(N-2)/2}P_2\nonumber\\
        & - \Bigavg{\bD_{\alpha1}^{UF}\cdot\nabla_1^T \ln P_{(N-2)/2}}_{(N-2)/2}P_2 - \bigavg{\bD_{\alpha1}^{UF}}_{(N-2)/2}\cdot\nabla_1^TP_2\nonumber \\
        &- \sum_{\beta=1}^2\Bigavg{\bD_{\alpha\beta}^{UL}\cdot\nabla_\beta^R \ln P_{(N-2)/2} }_{(N-2)/2}P_2 -\sum_{\beta=1}^2\bigavg{\bD_{\alpha\beta}^{UL}}_{(N-2)/2}\cdot\nabla_\beta^R P_2,
    \end{align}
    and
    \begin{align}
        \label{eq:jR-N-2}
        \int \bj_\alpha^R d \Gamma^{N-2}= & \bigavg{\bM_{\alpha1}^{\Omega F}}_{(N-2)/2}\cdot\bF_1^e P_2 - \Bigavg{\bD_{\alpha2}^{\Omega F}-\bD_{\alpha1}^{\Omega F} }_{(N-2)/2}\cdot\nabla_2^T P_2\nonumber \\
        & - \sum_{\beta=2}^N\Bigavg{\left(\bD_{\alpha\beta}^{\Omega F}-\bD_{\alpha1}^{\Omega F}\right)\cdot\nabla_\beta^T\ln P_{(N-2)/2} }_{(N-2)/2}P_2\nonumber\\
        & - \Bigavg{\bD_{\alpha1}^{\Omega F}\cdot\nabla_1^T \ln P_{(N-2)/2}}_{(N-2)/2}P_2 - \bigavg{\bD_{\alpha1}^{\Omega F}}_{(N-2)/2}\cdot\nabla_1^TP_2\nonumber \\
        & -\sum_{\beta=1}^2\Bigavg{\bD_{\alpha\beta}^{\Omega L}\cdot\nabla_\beta^R \ln P_{(N-2)/2} }_{(N-2)/2}P_2 -\sum_{\beta=1}^2\Bigavg{\bD_{\alpha\beta}^{\Omega L}}_{(N-2)/2}\cdot\nabla_\beta^R P_2 - D_\alpha^R \nabla_\alpha^R P_2.
    \end{align}
\end{widetext}
In equations \eqref{eq:jT-N-2} and \eqref{eq:jR-N-2}, we have defined $\avg{(\cdot)}_{(N-2)/2} = \int (\cdot) P_{(N-2)/2}d\Gamma^{N-2}$, and used the fact that the mobility tensors are independent of $\bq^N$ for spheres, i.e., $\bM_{\alpha\beta} = \bM_{\alpha\beta}(\br_2,...,\br_N)$.

In the dilute limit, neglecting the terms involving the gradients of $\ln P_{(N-2)/2}$ and using the pair mobility tensor in the absence of other particles in place of  $\avg{\bM}_{(N-2)/2}$, we obtain
\begin{align}
    \frac{\partial P_2}{\partial t} &+ \nabla_1^T\cdot \left(\bj_1^T - \bUtrap P_2\right) \nonumber\\
    &+ \nabla_2^T\cdot\left( \bj_2^T - \bj_1^T\right) +\sum_{\alpha=1}^2 \nabla_\alpha^R \cdot\bj_\alpha^R=0,
    \label{eq:pair-general}
\end{align}
where using the same symbols as before
\begin{eqnarray}
    \label{eq:jT-pair-general}
    \bj_\alpha^T &=& U_\alpha^0 \bq_\alpha P_2 + \bM_{\alpha1}^{UF}\cdot\bF_1^e P_2 - \left(\bD_{\alpha2}^{UF}-\bD_{\alpha1}^{UF} \right)\cdot\nabla_2^T P_2 \nonumber \\
    & & - \bD_{\alpha 1}^{UF}\cdot\nabla_1^T P_2 - \sum_{\beta=1}^2 \bD_{\alpha\beta}^{UL}\cdot\nabla_\beta^RP_2\\
    \label{eq:jR-pair-general}
    \bj_\alpha^R &=& \bM_{\alpha 1}^{\Omega F} \cdot\bF_1^e P_2 -  \left(\bD_{\alpha2}^{\Omega F}-\bD_{\alpha1}^{\Omega F} \right)\cdot\nabla_2^T P_2- \bD_{\alpha 1}^{\Omega F}\cdot\nabla_1^T P_2\nonumber \\
    &&  - \sum_{\beta=1}^2 \bD_{\alpha\beta}^{\Omega L}\cdot\nabla_\beta^RP_2- D_\alpha^R \nabla_\alpha^R P_2.
\end{eqnarray}

In the absence of hydrodynamic interactions, we have $\bM_{\alpha \beta}^{UF} =  \bI \delta_{\alpha\beta}/\zeta_\alpha^T$, $\bM_{\alpha\beta}^{\Omega L} = \bI \delta_{\alpha\beta}/\zeta_\alpha^R$,   and $\bM_{\alpha\beta}^{UL}, \bM_{\alpha\beta}^{\Omega F} = \bm{0}$, where $\delta_{\alpha \beta}$ is the Kronecker delta. The conditional probability of finding a bath particle, $\rho_{1/1}(\br_2, \bq_2, t \rvert \br_1, \bq_1, t)$, can be related to $P_2$ via the relation $\rho_{1/1} = (N-1)P_{1/1}$, where $P_{1/1}$ is defined by $P_2 = P_{1/1}P_1$. The factor of $N-1$ comes from the process of removing the ``labels'' of the $N-1$ bath particles. From this, the joint probability density of finding a bath particle at $\br_2$, $\bq_2$ and the probe at $\br_1$, $\bq_1$ is defined as $\rho_2 = \rho_{1/1}P_1$. Furthermore, we can define a dimensionless conditional distribution function $g_{1/1}$ such that
\begin{equation}
    \rho_2 = \rho_{1/1}P_1 = n_b g_{1/1}P_1,
\end{equation}
where $n_b = (N-1)/V$ is the number density of bath particles.

In the absence of hydrodynamic interactions, equations \eqref{eq:pair-general}-\eqref{eq:jR-pair-general} reduce to equations \eqref{eq:smoluchowski-pair-noHI}-\eqref{eq:jR-noHI} given in the text.

\section{Derivation of the variance relations}
\label{sec:appendix-pair-covariance}
For the pair problem, equation \eqref{eq:msd-PN} governing the mean-squared displacement reduces to
\begin{align}
    \label{eq:msd-pair}
    \frac{1}{2}\frac{\partial \avg{\br_1\br_1}}{\partial t}    +\frac{1}{\tau_k}\bigavg{\br_1\br_1}=& D_1^T \bI  +\left[U_1^0 \bigavg{\bq_1\br_1} - \bUtrap\avg{\br_1} \right]^\mathrm{sym} \nonumber \\ 
    &+D_1^T\Big[\int \nabla_2^T \rho_2 \br_1 d\Gamma^2 \Big]^\mathrm{sym}.
\end{align}

Using equations \eqref{eq:md-pair} and \eqref{eq:msd-pair}, one can show that the position fluctuation of the probe is governed by
\begin{align}
    \label{eq:var-r1-q1r1}
    \frac{1}{2}\frac{\partial \Var(\br_1)}{\partial t}  + \frac{1}{\tau_k} \Var(\br_1)= &D_1^T\bI+ U_1^0\left[\Cov(\bq_1, \br_1)\right]^\mathrm{sym} \nonumber \\
    &+D_1^T\Big[\int \nabla_2^T \rho_2 \Delta\br_1d\Gamma^2   \Big]^\mathrm{sym},
\end{align}
where $\Cov(\bq_1, \br_1) = \avg{\bq_1\br_1} - \avg{\bq_1}\avg{\br_1}$ and recall that $\Delta \br_1 = \br_1 - \avg{\br_1}$. To calculate the covariance of $\bq_1$ and $\br_1$ appearing in equation \eqref{eq:var-r1-q1r1}, we need $\avg{\bq_1\br_1}$,  $\avg{\bq_1}$ and $\avg{\br_1}$.

The net polar order of the probe satisfies
\begin{equation}
    \label{eq:avg-q1}
    \frac{\partial \avg{\bq_1}}{\partial t} +\frac{d-1}{\tau_1^R} \avg{\bq_1} = \bm{0},
\end{equation}
where $d (=2,3)$ is the dimensionality of the problem. It can be seen that the net polar order of the probe is not affected by the trap or the bath particles. The full solution to \eqref{eq:avg-q1} is readily obtained as
\begin{equation}
    \avg{\bq_1}(t) = \exp\left[-(d-1)t/\tau_1^R\right] \avg{\bq_1}(0),
\end{equation}
where any initial net polar order $ \avg{\bq_1}(0)$ decays away exponentially due to the rotary diffusion.

The average of $\bq_1\br_1$ is governed by
\begin{align}
    \label{eq:avg-q1r1}
    \frac{\partial \bigavg{\bq_1\br_1}}{\partial t} + \frac{1}{\tau}\bigavg{\bq_1\br_1}  =& -  \avg{\bq_1}\bUtrap + U_1^0 \bigavg{\bq_1\bq_1}\nonumber \\
    &+D_1^T \int \bq_1 \nabla_2^T \rho_2 d\Gamma^2  ,
\end{align}
where $\avg{\bq_1}$ is given by \eqref{eq:avg-q1} and $\bigavg{\bq_1\bq_1}$ satisfies
\begin{equation}
    \frac{\partial \bigavg{\bq_1\bq_1}}{\partial t} + \frac{2 d}{\tau_1^R} \left[\bigavg{\bq_1\bq_1} - \frac{1}{d}\bI\right]=\bm{0}.
\end{equation}
Similarly to $\avg{\bq_1}$, the net nematic order of the probe regardless of the presence of the trap or the bath particles is given by
\begin{equation}
    \label{eq:transient-qq}
    \avg{\bQ_1}(t) = \exp\left[-2dt/\tau_1^R\right]\bQ_1(0),
\end{equation}
where we have defined the net trace-free nematic tensor $\avg{\bQ_1} = \bigavg{\bq_1\bq_1} - \bI/d$.

At long times ($t\to \infty$), there is no net polar order of the probe, $\avg{\bq_1} = \bm{0}$ and the net nematic order is isotropic, $\bigavg{\bq_1\bq_1} = \bI/d$.

Using equations \eqref{eq:md-pair}, \eqref{eq:avg-q1} and \eqref{eq:avg-q1r1}, we obtain
\begin{align}
    \frac{\partial \Cov(\bq_1, \br_1)}{\partial t} +  \frac{1}{\tau}\Cov(\bq_1, \br_1) =& U_1^0\Cov(\bq_1,\bq_1) \nonumber\\
    &+ D_1^T\int \Delta\bq_1\nabla_2^T \rho_2d\Gamma^2 ,
\end{align}
where $\Delta \bq_1 = \bq_1 - \avg{\bq_1}$.

\section{Asymptotic analysis of the probe in the absence of bath particles}
\label{sec:appendix-transient-probe}
In equation \eqref{eq:covariance-q1-r1-single}, the timescale of transient decay $\tau$ can be written as 
\begin{equation}
    \frac{1}{\tau} = \frac{d-1}{\tau_1^R} + \frac{1}{\tau_k} = \frac{d-1+\epsilon}{\tau_1^R},
\end{equation}
where $\epsilon= \tau_1^R/\tau_k = k \tau_1^R/\zeta_1$. Using this definition, the solution of \eqref{eq:covariance-q1-r1-single} is given by
\begin{align}
    \Cov(\bq_1, \br_1)(t) = & e^{-t/\tau} \Cov(\bq_1, \br_1)(0) \nonumber \\ 
    &+ U_1^0 \int_0^t \exp \left(-\frac{t-s}{\tau}\right)\Var( \bq_1)(s)d s.
\end{align}
From equation \eqref{eq:transient-qq}, the preceding equation becomes
\begin{align}
    \Cov(\bq_1, \br_1)(t) =& e^{-t/\tau} \Cov(\bq_1, \br_1)(0) + \frac{U_1^0\tau \bI}{d} \left(1-e^{-t/\tau} \right) \nonumber \\
    & - \frac{\ell_1}{d+1-\epsilon}\left(e^{-2dt/\tau_R}- e^{-t/\tau}\right)\avg{\bQ_1}(0).
\end{align}
In the long-time limit ($t/\tau_R \gg 1$ and $t/\tau \gg 1$), we obtain equation \eqref{eq:covariance-q1-r1-single-result} in the text. Using equation \eqref{eq:var-r1-q1r1} in the absence of bath particles, we obtain
\begin{align}
    \label{eq:var-r1-single-transient-solution}
    \Var(\br_1)(t) =& e^{-2t/\tau_k}\Var(\br_1)(0)\nonumber\\
    & + \tau_k \left(D_1^T+ \frac{\left(U_1^0\right)^2\tau}{d}\right)\bI \left(1 - e^{-2t/\tau_k}\right)\nonumber \\
    & + 2 U_1^0 \int_0^t \exp\left[-2\frac{t-s}{\tau_k}\right]\left[\Cov^\prime(\bq_1, \br_1)(s)\right]^\mathrm{sym} d s,
\end{align}
where $\Cov^\prime(\bq_1, \br_1)(s) = \Cov(\bq_1, \br_1)(s) - U_1^0\tau \bI/d$ is the time-dependent (transient) part of the covariance of $\bq_1$ and $\br_1$. The integral in \eqref{eq:var-r1-single-transient-solution} can be carried out explicitly but is not important for the following discussion.

In the presence of the harmonic trap, the system exhibit two timescales that are important: the reorientation time $\tau_1^R$ and the viscoelastic timescale $\tau_k$; their relative importance is characterized by the parameter $\epsilon$. In the weak-trap limit, $\epsilon \to 0$, the two timescales are well-separated. It is useful to define the fast time variable $t_1 = t$ and the slow time variable $t_2 = \epsilon t$. We now consider the limit $\epsilon \to 0$ and the intermediate timescale in which the ABP has experienced many reorientations due to rotary diffusion but hasn't reached the ``boundary'' of the trap, i.e., $t_1/\tau_1^R \gg 1$ but $t_2/\tau_1^R = \epsilon t/\tau_R \ll 1$.

Differentiating equation \eqref{eq:var-r1-single-transient-solution} leads to
\begin{widetext}
    \begin{align}
        \label{eq:var-r1-single-transient-diff}
        \frac{d}{dt}\Var(\br_1)(t)= &\frac{-2}{\tau_k}e^{-2t/\tau_k}\Var(\br_1)(0) + 2 \left(D_1^T+ \frac{\left(U_1^0\right)^2\tau}{d}\right)\bI e^{-2t/\tau_k} \nonumber\\
        &+ 2U_1^0\left[\Cov^\prime(\bq_1, \br_1)(t)\right]^\mathrm{sym} 
        + 2U_1^0\frac{-2}{\tau_k} \int_0^t \exp\left[-2\frac{t-s}{\tau_k}\right]\left[\Cov^\prime(\bq_1, \br_1)(s)\right]^\mathrm{sym} d s.
    \end{align}
\end{widetext}
Since $1/\tau_k = \epsilon/\tau_1^R$ and
\begin{equation}
    \frac{t}{\tau_k} = \frac{t_2}{\epsilon \tau_k} = \frac{t_2}{\tau_1^R}\ll 1, \quad \frac{t}{\tau} = \frac{t_1}{\tau_1^R} (d+1-\epsilon)\gg 1,
\end{equation}
we have
\begin{align}
&e^{-2t /\tau_k}  = e^{-2 t_2/\tau_1^R} = 1+ O(t_2/\tau_1^R),\\
&\frac{\left(U_1^0\right)^2\tau}{d} = \frac{U_1^0\ell_1}{d(d-1)}\left[ 1 +O(\epsilon)\right].
\end{align}
Therefore, equation \eqref{eq:var-r1-single-transient-diff} at leading order is
\begin{align}
    \frac{1}{2}\frac{d}{dt}\Var(\br_1)(t) &=  \left(D_1^T +  \frac{U_1^0\ell_1}{d(d-1)}\right)\bI \nonumber\\
    &= \left(D_1^T+D_1^\mathrm{swim}\right)\bI.
\end{align}
It is clear that in the weak-trap limit in this intermediate timescale, the ABP exhibits a diffusive behavior with the free-space diffusivity $D_1^T+D_1^\mathrm{swim}$.

We now consider the weak-trap limit but at long-times, $t/\tau_1^R\gg 1, t/\tau_k \gg 1$. So long as the trap strength is not identically zero, the ABP will eventually ($t/\tau_k \gg 1$) experience the confinement of the trap. Using equation \eqref{eq:var-r1-single-transient-solution}, we obtain at long times
\begin{equation}
   \frac{1}{\tau_k} \Cov(\br_1, \br_1)\to \left(D_1^T+D_1^\mathrm{swim}\right)\bI.
\end{equation}

In the strong-trap limit ($\epsilon \to \infty$) and at long times, we have $\tau/\tau_1^R = O(1/\epsilon) $ and the position fluctuation of the probe $\Var(\br_1)/\tau_k =  D_1^T$.

\end{document}